\documentclass{article}

\usepackage{xurl}
\usepackage{listings}
\usepackage{placeins}

\usepackage{microtype}
\usepackage{graphicx}
\usepackage{booktabs} % for professional tables

\usepackage{hyperref}

\usepackage{epsfig,endnotes}
\usepackage{subcaption}
\usepackage{xspace}
\usepackage{balance}

\usepackage{fancyhdr}
\usepackage{extramarks}
\usepackage{amsmath}
\usepackage{amsthm}
\usepackage{amsfonts}
\usepackage{tikz}
\usepackage[accepted]{mlsys2021}
\mlsystitlerunning{A Tale of Two Models: Constructing Evasive Attacks on Edge Models}

\newcommand{\eg}{e.g.,\xspace}
\newcommand{\ie}{i.e.,\xspace}

\DeclareMathOperator*{\argmax}{arg\,max}

\DeclareMathOperator{\sign}{sign}
\DeclareMathOperator{\model}{model}
\DeclareMathOperator{\Clip}{Clip}

\newcommand{\name}{DIVA\xspace}

\newcommand{\resnet}{ResNet}
\newcommand{\mobilenet}{MobileNet\xspace}
\newcommand{\densenet}{DenseNet}

\newcommand{\whitebox}{whitebox\xspace}
\newcommand{\blackbox}{blackbox\xspace}
\newcommand{\orig}{original\xspace}
\newcommand{\fp}{full-precision\xspace}
\newcommand{\q}{adapted\xspace}
\newcommand{\finetune}{finetune\xspace}

\usepackage{todonotes}

\usepackage{enumitem}
\setitemize{noitemsep,topsep=0pt,parsep=0pt,partopsep=0pt}
\setenumerate{noitemsep,topsep=0pt,parsep=0pt,partopsep=0pt}
\setdescription{itemsep=0pt,topsep=0pt,parsep=0pt,partopsep=0pt,itemindent=0em,leftmargin=0.5em}

\newcommand{\paragraphb}[1]{\noindent{\bf #1}}

\newenvironment{denseenum}{
\begin{enumerate}[topsep=2pt, partopsep=0pt, leftmargin=1.5em]
  \setlength{\itemsep}{2pt}
  \setlength{\parskip}{0pt}
  \setlength{\parsep}{0pt}
}{\end{enumerate}}

\begin{document}

\twocolumn[
\mlsystitle{A Tale of Two Models \\\large Constructing Evasive Attacks on Edge Models}

% unless you've provided the [accepted] option to the mlsys2020


\mlsyssetsymbol{equal}{*}
\mlsyssetsymbol{notmeta}{\dag}

\begin{mlsysauthorlist}
\mlsysauthor{Wei Hao}{columbia}
\mlsysauthor{Aahil Awatramani}{cornell}
\mlsysauthor{Jiayang Hu}{columbia}
\mlsysauthor{Chengzhi Mao}{columbia}
\mlsysauthor{Pin-Chun Chen}{columbia}
\mlsysauthor{Eyal Cidon}{meta,notmeta}
\mlsysauthor{Asaf Cidon}{columbia}
\mlsysauthor{Junfeng Yang}{columbia}
\end{mlsysauthorlist}

\mlsysaffiliation{columbia}{Columbia University}
\mlsysaffiliation{cornell}{Cornell University}
\mlsysaffiliation{meta}{Meta \*\dag This research was done independently of Meta}
\mlsyscorrespondingauthor{Wei Hao}{wh2473@columbia.edu}


\mlsyskeywords{Machine Learning, MLSys}

\vskip 0.3in

\begin{abstract}
%\subsection*{Abstract}
%
Full-precision deep learning models are typically too large or costly to deploy on edge devices. To accommodate to the limited hardware resources, models are adapted to the edge using various edge-adaptation techniques, such as quantization and pruning. % Models running on the edge are commonly \emph{quantized}, which means they use fewer bits to represent numbers when performing computations and storing data. 
While such techniques may have a negligible impact on top-line accuracy, the adapted models exhibit subtle differences in output compared to the \orig model from which they are derived.
%and the original model on adversarial examples are not thoroughly investigated. \cz{modified, we are the first one to notice this difference.}
 %  the adapted models exhibit subtle \cz{severe} differences compared to the original model when run on particular inputs.
In this paper, we introduce a new evasive attack, \name, that exploits these differences in edge adaptation, by adding adversarial noise to input data that maximizes the output difference between the original and adapted model. 
Such an attack is particularly dangerous, because the malicious input will trick the adapted model running on the edge, but will be virtually \emph{undetectable} by the original model, which typically serves as the authoritative model version, used for validation, debugging and retraining. %.Since it is easier for an attacker to obtain access to the model running on the edge device, we show how \name can be adapted to a semi-\blackbox scenario where the attacker has access only to the \q model. 
We compare \name to a state-of-the-art attack, PGD, and show that \name is only 1.7--3.6\% worse on attacking the adapted model but 1.9--4.2$\times$ more likely not to be detected by the the \orig model under a whitebox and semi-\blackbox setting, compared to PGD.
%We evaluate several existing and novel defense mechanisms against \name, but find that all of them are not effective in stopping it. %We suspect that it is fundamentally difficult to defend against such an attack, due to the inherent differences between the original and adapted models. 
%Finally, we demonstrate how \name can be used to attack a face recognition model running on an edge device (\eg a security camera) with a success rate of 97.5\%. Notably, \name causes the model to misidentify Nicolas Cage as Jerry Seinfeld!
\end{abstract}
]

\printAffiliationsAndNotice{} % otherwise use the standard text.

\section{Introduction}
 
Deep learning (DL) models are increasingly being deployed in large-scale applications on millions of edge devices, such as phones and cameras. Notable real-world edge deployments include language translation~\cite{translate-example}, object detection~\cite{object-example}, face recognition~\cite{fd-example} and ad recommendations~\cite{wu2019machine}.
The common lifecycle of edge-distributed DL models is to assemble a large dataset, and train the model on a powerful set of servers, using DL accelerators (\eg GPUs, TPUs). After the model has been trained, it is typically pushed to a set of edge devices, where it will conduct inference~\cite{mistify,tflite,wu2019machine}. 
%a different set of devices, where it is used for inference. The devices used for inference may be very different than the ones used for training: for example, computer vision (CV) models might conduct inference on a mobile phone, or on the camera of a drone or a self-driving car. 
Edge devices are resource constrained compared to the servers used to train the model: they have weaker accelerators, less DRAM, limited power and energy constraints. Therefore, the model needs to be \emph{adapted} to run on these devices.
%\newtodo{J: not sure we need the following.} %Inevitably, in production environments, due to model drift~\cite{gama2014survey}, models need to constantly retrained and validated. Since training is much more expensive than inference, the retraining and validation are also typically done on powerful servers\footnote{Federated learning, a technique for training models from a distributed set of edge devices, is still not the commonly-used approach.}.

To adapt models to resource-constrained edge devices, several techniques are commonly used, including quantization, distillation and pruning~\cite{deng2020model}. Such techniques shrink the size and representation of the \orig model. For example, quantization converts the floating-point representation of the original model to a more compact form, such as int8 or int4, which reduces the computational complexity, the amount of memory, and the energy spent when conducting inference. %The level of quantization (\ie number of bits) is determined by the edge device's constraints. 
While quantization generally reduces the accuracy of the model, on average, its effect is relatively modest;
%. In a set of experiments on three standard CV architectures (\resnet50, \mobilenet and \densenet121), 
in our experiments the the \q int8 version achieves at least 96\% of the accuracy of the \orig model on average across three standard computer vision models.
However, despite the small differences in top-line accuracy, the models are not identical, and may return slightly different results for a particular input. %To characterize these variances in output, we compared the predictions a \orig model and an int8 \q one, on three architectures with a set of 30,000 images, 
%On ImageNet, we find that up to 8.1\% of images are predicted correctly by one model but mispredicted by the other. Beyond the final prediction, both models have different confidence scores even when they are both correct. 
% Asaf: we don't need to talk about pruning here.
%\todo{for pruning, the pruned model's accuracy for mobilenet is 62.1\% of its original model but its reasonable since its structure is already highly optimized and can not afford pruning. Should we talk about pruning here?}

Our key observation is that \emph{subtle differences between the edge-adapted and the \orig models can be exploited by an attacker}. 
We propose \name (DIfferential eVasive Attack), an attack against computer vision models, which causes the edge-adapted model to mispredict, while remaining virtually undetectable to the \orig model. 
In large-scale production environments, ML operators run thousands of slightly-different adapted models on tens of thousands of device models~\cite{wu2019machine}. Such an attack would make it very difficult for the operator to detect and debug, because when validating the input against their authoritative \orig model they would not detect the input as malicious. Therefore, the operator may assume that no attack occurred until the attackers have subverted a significant portion of the edge devices. Even if the operator detected it, it would be expensive and time-consuming to debug the root cause and understand which edge models it affects.

\name constructs an attack image (indistinguishable to a human from the original image), with the goal of maximizing the prediction loss of the adapted model, while minimizing the prediction loss of the original model. \name uses simple iterative optimization methods, such as stochastic gradient descent, to efficiently construct the attack image. This allows \name to create an input that would cause the \q model to mispredict, while causing minimal changes to the prediction of the original model. 
We initially design \name assuming a \whitebox setting, where the attacker has access both to the \orig and the adapted model. We leverage differential testing~\cite{mckeeman1998differential,pei2017deepxplore}, a powerful technique for detecting deviations of two implementations of the same functionality to generate adversarial samples. Compared to searching through a vast space of possible behaviors of one model, \name focuses on a much smaller space of deviations of two very similar models, and is therefore very effective at generating successful attacks.
We show \name generalizes across two popular edge-adaptation techniques: quantization and pruning.

%Since the version of the model that is typically used for validation, debugging, and retraining is the \orig one, \name would be undetected by most standard machine learning pipelines, including those used by popular machine learning frameworks (\eg Tensorflow~\cite{tensorflow}, Tensorflow Lite~\cite{tflite}, and PyTorch~\cite{pytorch-quantization}). %, which by default conduct retraining and validation on the base, \fp model.

Since it may be much easier for the attacker to obtain access to the \q model, which could be running on any edge device (\eg phone, camera), rather than the \orig model, which might be stored in a more secure location (\eg in the cloud), we also create a semi-\blackbox attack for quantized models that requires only the \q model. To this end, we reconstruct a \fp surrogate model that shares the same architecture with the \q model, by teaching the surrogate model to imitate the \q model via knowledge distillation~\cite{BaC14,PolinoPA18}. We then generate the adversarial attack on this surrogate model along with the \q model and evaluate its efficacy on the original \fp and the \q models. Finally, we test \name in a blackbox setting where the hacker does not have access to the parameters of the \q model as well and tries to reconstruct both a surrogate \fp and a surrogate \q model via knowledge distillation followed by adaptation.

Our evaluation shows that both \whitebox and semi-\blackbox \name significantly outperform a state-of-the-art attack, PGD~\cite{PGD}, in causing the \q model to mispredict while not affecting the \orig model. The \whitebox \name attack is able to cause the \q model to mispredict and \orig model to predict correctly 92.3--97\% for quantization and 98--100\% for pruning, across three architectures (\resnet, \mobilenet, \densenet) on ImageNet, while the semi-\blackbox \name attack does so with a success rate of 71.1--96.2\%. Even fully blackbox \name outperforms PGD, albeit with a lower success rate of 30.3--77.2\%.
Furthermore, despite the constraint not to affect the \orig model, \name is only up to 3.6\% worse than PGD in misleading the edge models. Therefore, from the attacker's standpoint, there is very little cost in using \name compared to an attack that targets solely edge models.

% We further show that \name is effective even if the model is equipped with the state of-the-art robust defenses. We adapt several robustness training methods to our scenario, including using min-max robust training specifically tailored against \name and applying knowledge distillation during quantization-aware-training, and different variations and combinations of these two. We show that none of the defenses is very effective against \name. Most of the robustness methods we evaluate in fact make the model even more succeptible to attack, and the only one that is able to reduce its effectiveness is able to do so by only 4.9\%. 
% We believe the fundamental reason that \name is so powerful is that, like prior successful differential testing tools for software vulnerabilities~\cite{chen2015guided,mckeeman1998differential}, \name can quickly zoom into the small space of deviations of the two models. Since the \orig and the \q models will always have subtle differences due to model life cycle requirements, we believe attacks that exploit the difference between two models are fundamentally difficult to defend against. Given the widespread prevalence of model adaptation to edge devices, mitigating such attacks presents an important direction of future research for the community.

Finally, to showcase a real-world attack scenario, we demonstrate how \name can be used to attack a face recognition model that would be running on a security camera or a phone. Similar to the ImageNet experiment, the whitebox \name attack achieves a 98\% success rate here. % with the face recognition model. %Among the amusing adversarial samples that it generates, one example causes the \q model to misidentify Nicolas Cage as Jerry Seinfeld but results in a correct prediction on the \orig model. 
%\todo{Maybe we don't want to show the result of targeted attack here since it is not that great?}

We make the following contributions:
\vspace{-0.5em}
\begin{denseenum}
    \item We present a new vulnerability introduced by the widespread use of model adaptation, which causes slight variations between the models trained on the server and conducting inference on the edge. 
    \item \name is a new attack that targets \q computer vision models running on edge devices, which is virtually-undetected by validation on the \orig model. We show that adding this evasive property to adverserial attacks comes at a low cost for the attacker. %A similar approach can be used to attack other edge-adaptation techniques, such as compression and pruning.
    \item \name is effective even in a semi-\blackbox setting where the attacker can only access the \q model. We design a novel attack where the attacker tries to reconstruct the \orig model using distillation learning, and is able to successfully attack the \q model relatively undetected. 
    % \item we design novel defenses for \name, and show that they are ineffective, since \name is powerful at exploiting the differences between two models.
    \item We show \name can be used in a realistic scenario of a face recognition model deployed on an edge device.
\end{denseenum}

\section{Background and Closely Related Work}
\label{sec:background}

This section presents a primer on model quantization, pruning and distillation (\S\ref{sec:quantization}), surveys the latest adversarial attacks against DL models (\S\ref{sec:attacks}) and defenses (\S\ref{sec:robust}), and introduces differential testing and its application to DL (\S\ref{sec:differential}), which uses techniques relevant to our work.

\subsection{Quantization, Pruning and  Distillation} \label{sec:quantization}

\paragraphb{Basic quantization.}
DL models are trained using a floating point representation with 32 or 64 bits~\cite{fp-precision,nvidia-fp}. These floating point representations are typically supported by server-based accelerators~\cite{nvidia-fp,cerebras-fp}. 
When deployed to edge devices, models are often quantized in order to reduce their resource consumption. Quantization reduces the size of the model and its resource consumption when conducting inference, by representing the model using a smaller number of bits for each number (\eg 8 bits instead of 32). This typically requires changing the representation from a floating point to an integer. While quantization reduces the compute, memory, and energy footprint, as well as the latency of inference~\cite{DBLP:conf/cvpr/CaiHSV17,DBLP:journals/corr/HanMD15,DBLP:conf/icml/LinTA16}, it also reduces the model's accuracy. 

\paragraphb{Quantization-aware training.}
%Since quantization reduces the number of bits used to represent a number, the model is unable to represent the same range of information, which causes reduced accuracy. 
Quantization-aware training (QAT)~\cite{JacobKCZTHAK18} is a technique that improves the accuracy of the \q model by introducing quantization noise \emph{during the training of the model}. In QAT the base model training is still done in full precision. During the forward pass, weights and data are \q to a lower precision (\eg from float32 to int8) and increased back to the original precision, in the backpropagation the gradients are applied to the weights via straight-through estimators \cite{DBLP:journals/corr/BengioLC13}. This forces the base model to learn weights that are more robust to the quantization error.

\paragraphb{Pruning.}
Pruning is a technique that zeros out model weights that are insignificant and barely used during the training process to achieve sparsity. A sparse model has the advantages of easier compression, smaller model size and higher inference speed, resulting in a smaller and faster neural network. This technique is being used for vision and translation models, and is being evaluated in other scenarios, such as speech applications ~\cite{pruneZhuG18}.

\paragraphb{Model distillation.}
Another common technique to reduce model size, is model distillation~\cite{BaC14,hinton2015distilling,PolinoPA18}. Distillation reduces the model size by training a smaller model (student model) that learns to match the output of a larger model (teacher model), given the same inputs. After training, the student model captures almost the same information as the teacher model but with fewer parameters. 
% Distillation can be used not only to reduce the number of parameters in a model but also to train a better-quantized model~\cite{PolinoPA18} \todo{Distillation can be used not only to reduce the number of parameters in a model but also to train a better-quantized model. maybe we don't need this sentence?}.

\subsection{Adversarial Attacks Against DL Models} \label{sec:attacks}

A plethora of research is devoted to creating adversarial attacks against DL models~\cite{shan2020gotta,blind-adverserial,SLAP,Waveguard,policy-training,metric:neurips19,gu2014towards,goodfellow2014explaining,carlini2017towards,distillation-defense,carlini-wagner-adverserial,papernot2016limitations,R_plus_FGSM}. The basic idea is to find a small, human-imperceptible perturbation under some bound (e.g., $\ell_\infty$-norm), such that when the perturbation is added to the input sample, the model mispredicts. Depending on the underlying threat model, adversarial attacks can be \whitebox (attackers have full access to model parameters) or \blackbox (attackers have no access to model parameters). 

\paragraphb{Whitebox attacks.} In a \whitebox attack setting, attackers have access to the model's parameters. An attacker can use those to create a perturbation that maximizes the following loss function for a particular input and label:
\vspace{-0.2em}
\begin{equation}\label{eq:attack}
    A = \argmax_{A \leq \epsilon} L(\theta, x + A, y)
\end{equation}
\vspace{-0.1em}
\noindent
\vspace{-0.3em}
where $L$ is the loss function, $\theta$ the model parameters, and $A$ an adversarial perturbation constructed for input $x$ with label $y$ under perturbation bound $A \leq \epsilon$.

While Equation~\ref{eq:attack} can be solved using an expensive method (box-constrained L-BFGS)~\cite{DBLP:journals/corr/SzegedyZSBEGF13}, a much more efficient method with linear complexity is the commonly-used Fast Gradient Sign Method (FGSM)~\cite{goodfellow2014explaining}, in which an input is distorted by adding small perturbations $\epsilon$ with opposite signs to the gradients. By making the signs of the input perturbation opposite to the gradient, each dimension of the input will enlarge the error by $\epsilon$, and a high-dimensional input leads to a large accumulative error. Specifically, FGSM generates an adversarial sample $\hat{x}$ with ($\nabla_{x}$ is derivative to $x$):
\begin{equation}
    \hat{x}=x+\epsilon*\sign\left(\nabla_{x}L\left(\theta,x,y\right)\right)
\end{equation}

R+FGSM~\cite{R_plus_FGSM} is an enhancement to FGSM that adds random perturbations to the noise in addition to the gradient maximization. Projected Gradient Descent (PGD)~\cite{PGD} further improves the attack by iteratively maximizing the loss
\begin{equation}\label{eq:pgd}
    \hat{x}_{t+1} = \Clip_{x, \epsilon}\{ \hat{x}_{t} + \alpha \sign(\nabla_{x}L( \theta, x, y)) \}
\end{equation}

\noindent
where Clip$_{x, \epsilon}$ is a projection function to the region defined by input $x$ and perturbation bound $\epsilon$ in the input space and initially $\hat{x}_0 = x$. The idea in PGD is to convert the one-step FGSM attack into multiple steps. In step $t+1$, FGSM is applied to step $t$'s adversarial sample $\hat{x}_t$ with a step perturbation size of $\alpha$, and the result is projected back to the overall perturbation bound $\epsilon$. PGD is thus determined by three parameters: the number of steps $t$, the step size $\alpha$, and the perturbation strength $\epsilon$. PGD is one of the state-of-the-art adversarial attacks, and we use it as the primary baseline in our evaluation (\S\ref{sec:eval}). We also evaluate other baseline methods: Momentum PGD \cite{PDGm} and CW \cite{carlini2017towards} attacks, which perform worse than PGD (\S\ref{sec:otherbaselines}).

\paragraphb{Blackbox attacks.} Unlike \whitebox attacks which allow full access to the model parameters in order to calculate gradients, \blackbox attacks allow no such access. Prior work has demonstrated successful attacks in \blackbox settings by learning a surrogate model and then attacking it instead~\cite{papernot2017practical}. The insight is that adversarial attacks target fundamental weaknesses in the learning method, and are thus transferable from one model to another~\cite{papernot2016transferability}.

\paragraphb{Bit-Flip attack.}
Another form of attack is altering the model in memory. The bit flip attack \cite{rakin2019bit} is a general parameter attack that targets the weights of quantized DNNs by bit-flipping their values in memory, based on the idea that resource-limited platforms normally lack effective data integrity check mechanism. This attack is largely orthogonal to the aforementioned attacks: it makes changes to the device’s memory, while the later makes changes to its input during the attack phase. Such an attack can be mitigated by performing checksums. Moreover, attacks on inputs can be targeted to a large group of devices, as they do not require device infiltration.  

\paragraphb{Attacks against \q models.} Prior work on Defensive Quantization applies the PGD attack to both the \orig and the \q models~\cite{defensive_q} and shows that with a small perturbation bound, \q models are more robust than their \orig counterparts. However, once the perturbation increases beyond a certain point, \q models exhibit larger errors and are less robust.

%\paragraphb{Discussion}
%Almost all prior adversarial attacks target one model, in contrast to our work, which targets the deviation of the \q model from the \orig version. Besides whitebox mode, \name can also be carried out in semi-blackbox mode, in which attackers have full access to the \q model but not the \orig version, because it is conceivably much easier to obtain the \q model from an edge device than hacking into a training server. \newtodo{Consider moving this paragraphb to the end of this section or even later. Right now we haven't introduced \name, so this lacks context}

\subsection{Robust Training} \label{sec:robust}

Much recent work defends against adversarial attacks via robust training~\cite{han2018co,hendrycks2019using,wang2019symmetric,wong2020fast}. Robust training is typically formulated as a minimax problem and optimizes the model parameters to minimize the maximum loss an attack can induce~\cite{PGD,lanckriet2002robust}:
\vspace{-0.1em}
\begin{equation}\label{eq:robust}
    \min_{\theta}\,\max_{A \leq \epsilon}L(\theta, x + A, y)
\end{equation}
\vspace{-0.1em}
\noindent
This minimax method is considered a principled approach with roots from robust optimization~\cite{scarf1958min}. It has been shown to boost model accuracy under attack by 28.6\% \cite{DBLP:journals/corr/abs-2103-10282}, achieving the state of the art.
Robust training is more expensive than regular training because of the high cost of solving the minimax as opposed to the minimization problem. Therefore it is typically only applied to the \orig \fp models on powerful servers.

\subsection{Differential Testing} \label{sec:differential}

Our work on exploiting the differences between two versions of a model is inspired by differential testing, a popular software testing method that detects the deviations between two implementations of the same functionality. It feeds the same input to the two implementations, observes the differences in the executions, and attempts to mutate the input to cause to larger differences. It has been shown to effectively detect many semantic bugs including SSL certification verification vulnerabilities~\cite{chen2015guided}.
The power of differential testing lies in that it cross-checks two implementations, using each as the reference for the other. Compared to the vast space of possible execution behaviors of one implementation, the deviations of the two implementations form a much smaller space. %Differential testing is tailored to guide the search of bugs to this small deviation space, a key characteristic that enables it to perform so well in software vulnerability detection.

A few prior systems applied differential testing to DL models~\cite{pei2017deepxplore,guo2018dlfuzz,guo2020coverage}. For instance, DeepXplore~\cite{pei2017deepxplore} generates adversarial samples with physically-realizable transformations (e.g., brightening and rotation) by finding deviations of two DL models. It formulates the problem as a joint optimization with several constraints, including that the perturbations must map to one of the transformations, and that the number of activated neurons is maximized. Like these systems, we employ the idea of differential testing but apply it to a different goal of differentially-attacking the \q model, while referencing the \orig model. %As such, \name solves a different joint optimization problem, elaborated in \S\ref{sec:dqa}.
\section{Motivation}
\label{sec:motivation}

In this section we provide further motivation for \name. We describe the attack scenario, and then motivate \name using the example of quantization, a common edge-adaptation technique. We demonstrate that while quantization has a small impact on average accuracy, it causes bigger deviations on individual inputs. Finally, we demonstrate that existing attacks against quantized models also significantly affect the prediction of the \orig model, and thus would be easier to detect and debug when the \orig model undergoes validation and debugging.
%existing attacks may affect \q models more than the full-preci existing attacks against DL models. We show how these attacks can be used to effect \q models, which are more vulnerable than \orig models to adversarial attacks~\cite{defensive_q}.
%\name is an attack which leverages the divergence between the \q model and the \orig model, we explain what is the benifits of using \name relative to other adversarial attacks. \newtodo{(Eyal:) maybe briefly introducing \name here helps with contextualizing Figure\ref{fig:pca-analysis} and \ref{fig:pgd_vs_dqa}}

\paragraphb{Attack scenario.}
Companies that deploy models on edge devices need to deal with the huge diversity in hardware on different phones, tablets and cameras. For example, Facebook estimates that devices running its deep learning models comprise over 2000 unique SoCs running on tens of thousands of tablet and phone models. To accommodate such a large set of devices, for each one of their \fp models, ML operators may need to create thousands of edge-adapted models versions~\cite{wu2019machine}. Even worse, since devices come online sporadically, some of the devices may still be running older versions of the adapted models. Due to the very large number of adapted models, and their similarity in top-line accuracy to the \orig model, ML operators typically only use the \orig version of the model for validation and debugging, which is also the case in popular machine learning frameworks~\cite{tensorflow,tflite,pytorch-quantization}. 

Therefore, an attack that only targets a particular edge model, but not the \orig model, would be harder to detect by the operator, and even if it was detected, it would be expensive to track down and debug. For example, in order to understand which devices and which model versions may be vulnerable to a particular set of adversarial inputs, the ML operator might have to sift through many adapted models or, worse, re-generate all the adapted models for all model versions on all possible edge devices, and manually run inference for each one of these models on the adversarial inputs. This process is time-consuming and expensive, and before the operator pinpoints the root cause of the attack, many edge devices may have been subverted.

We now provide motivation for how to construct such attacks using the example of quantization.

\begin{table*}[!t]
\centering
\small
\begin{tabular}{|l|r|r|r|r|r|}
\hline
& \multicolumn{1}{|l|}{Original} & \multicolumn{1}{|l|}{Quantized} & \multicolumn{1}{|l|}{Original Correct \&} & \multicolumn{1}{|l|}{Original Incorrect \&}  & \multicolumn{1}{|l|}{Total} \\
Architecture & \multicolumn{1}{|l|}{Accuracy} & \multicolumn{1}{|l|}{Accuracy} & \multicolumn{1}{|l|}{Quantized Incorrect} & \multicolumn{1}{|l|}{Quantized Correct} & \multicolumn{1}{|l|}{Instability} \\
\hline
\hline
\resnet50 & 72.1\% & 70.1\% & 1510 & 925 & 8.1\%\\
\hline 
\mobilenet & 69.1\% & 67.4\% & 1199 & 677 & 6.3\%\\
\hline
\densenet121 & 73.5\% & 71.0\% & 1567 & 816 & 7.9\%\\
\hline
\end{tabular}
\caption{Comparing accuracy between the \orig \fp models and quantized models, and the number of samples that deviate in both models. Instability is the total percentage of images that deviate between the models.}
\vspace{-1em}
\label{tab:avg-accuracy}
\end{table*}

\paragraphb{The impact of quantization on accuracy.}
In general, quantization has a relatively small impact on the \emph{average accuracy}. Table~\ref{tab:avg-accuracy} compares the average accuracy of the \orig floating-point model (fp32) to a model that uses quantization-aware training (int8 precision) across three different architectures. Across all three architectures, the accuracy achieved by the quantized model is 96\% or more of the original model's accuracy.

While quantization does not significantly affect the top-line average accuracy, it introduces a subtle variance in prediction for individual inputs. The table shows the number of deviations in predictions, where one model predicts correctly and the other incorrectly. We also calculate the instability~\cite{cidon2021characterizing} across the two versions -- the percentage of total images where the models disagree. Across all three architectures, the instability varies between 6.3--8.1\%, which is higher than the top-line accuracy metric on its own would suggest.

\begin{figure}
\small
\centering
\includegraphics[width=0.35\textwidth]{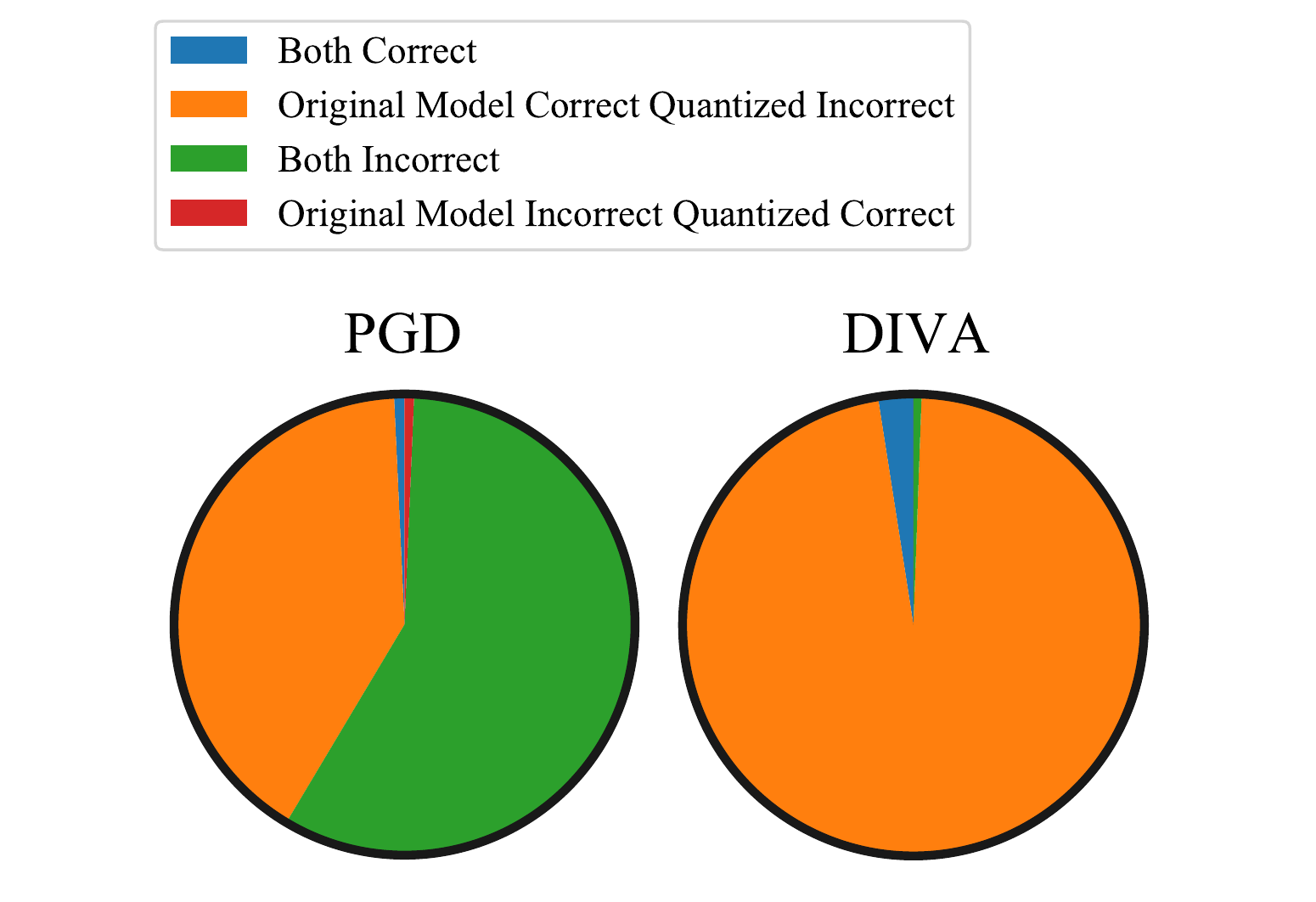}
\caption{Results of the PGD and \name attacks on the \orig \fp \resnet50 and its quantized version. PGD causes both versions of the model to misclassify. On the other hand, \name causes a misclassification only on the \q version and mostly does not affect the \orig version. }
\label{fig:pgd_vs_dqa}
\end{figure}

\paragraphb{Baseline causes \fp models to mispredict.} %Prior work has shown that \q models are more vulnerable to adversarial attacks then their \orig versions~\cite{defensive_q}. 
%The same methods discussed in \S\ref{sec:attacks} can be used against a quantized, \whitebox model, and they will typically be even more successful in causing it to misclassify. 
Adversarial attacks have been shown to be transferable from one model to another~\cite{papernot2016transferability}. In other words, an attack that affects one model is likely to impact another. This means that if an attacker targets a quantized model using a standard attack, it is likely to also affect the \orig model. 
Indeed, Figure~\ref{fig:pgd_vs_dqa} shows the percent of images that get classified correctly and incorrectly by the \orig model and the quantized one when we apply PGD on a quantized \resnet50. The results show that PGD also effects the \orig model. Thus, since validation is typically conducted on the \orig model, it is likelier that such an attack would be detected. On the other hand, we will show that \name (\S\ref{sec:dqa}), successfully attacks the quantized model, but it has a much smaller effect on the \orig model, and so it is likely to go undetected during validation. 

\section{Differential Evasive Attack} \label{sec:dqa}

This section describes the construction of \name, starting from the intuition (\S\ref{sec:idea}), followed by the whitebox (\S\ref{sec:whitebox}), semi-\blackbox (\S\ref{sec:semi-blackbox}) 
 and \blackbox attacks (\S\ref{sec:blackbox}).

% In this section we introduce \name. We start by discussing a whitebox scenario, where the attacker has access to \emph{both} the \orig and \q models (\S\ref{sec:whitebox}), and their goal is to maximize the loss to the \q model, while minimizing the loss to the \orig one. Then, we discuss how \name can be used in a semi-\blackbox scenario, where the attacker only has access to the \q model, but not the original \orig one (\S\ref{sec:semi-blackbox}). In this scenario, the attacker cannot precisely compute the loss of the \orig model. 

%\begin{equation}\label{pgd}
%    x_0^{adv} = x,
%    x_{N+1}^{adv} = \pi_{x, \epsilon}\{ x_{N}^{adv} + \alpha sign(\nabla_{x}L( %    \theta, x_{N}^{adv}, y)) \}
%\end{equation}

% Equation \ref{pgd} describes the attack where A is the generated attack noise, image is the original input, $\theta$ is the model parameters. We do not initial the attack using random noise because random start is less effective in single run. For PGD attack, the attackers have access both to the \q model and the \orig model.

\subsection{Intuition for \name} \label{sec:idea}

\name targets the deviation of the \q model from the \orig model. The goal of \name is to be stealthy, because the adversarial samples that it generates mislead only the \q model but otherwise behave correctly on the \orig model, making it challenging for the machine learning engineer to detect the attack during training and validation, which is typically conducted on a server or cluster setting. Even when detected, \name exposes vulnerabilities that are sticky and difficult to debug and remedy, because virtually all existing robust training techniques target \orig models and cannot directly apply to \q models. While robust training may conceivably be 
accommodated to work with model adaptation, the model parameters are eventually compressed, ``coarsening'' the decision boundaries in the input space. The coarse-grained decision boundaries of the \q model always deviate slightly from the fine-grained decision boundaries of the \orig model. However subtle the deviations may be, \name is designed to effectively locate them and exploit them to launch successful attacks. %We believe the \name approach will be effective at attacking other edge-adaption techniques such as compression and pruning because, by design, these techniques create deployment models with subtle differences from the original.

\begin{figure}
    \centering
    \includegraphics[width=0.25\textwidth]{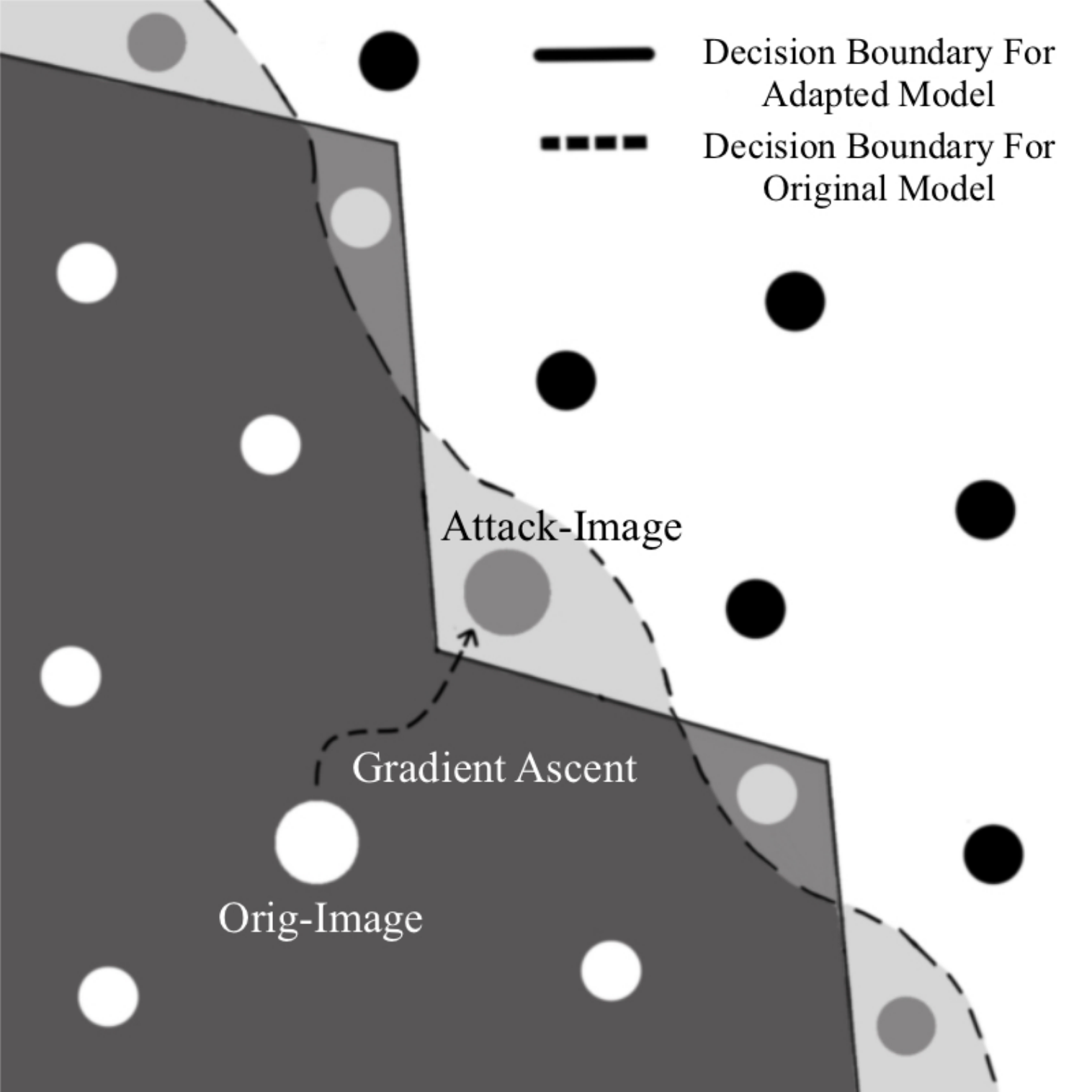}
    \caption{%Illustration of how \name finds the subtle deviations between the \orig and \q models. 
    The decision boundaries of the \q model in the input space are coarser-grained than those of the \orig model. \name leverages differential testing to locate the small deviations along the boundaries.}
    \vspace{-1.2em}
    \label{fig:idea}
\end{figure}

Figure~\ref{fig:idea} shows how \name uses differential testing to locate the subtle deviations. Given a sample, \name observes how each of the two models compute the label for the sample. It then calculates a small perturbation that keeps the \orig model's label unchanged but reduces the probability for the \q model to predict the same label. \name can jointly consider both models' probabilities and calculate the input perturbation accordingly. It repeats this step multiple times until it reaches one of the regions in the input space where the \q and \orig models deviate.

\subsection{Whitebox \name Attack} \label{sec:whitebox}

% The whitebox attacker's goal is to maximize the loss for the \q model, while minimizing the loss for the \orig model, so the attack can go undetected when the model is validated and debugged, which is typically done on the \orig model.

In the \whitebox \name attack, the attacker has full access to both the \orig and \q models. Similar to the general \whitebox adversarial attack, which we described in Equation~\ref{eq:attack}, the attacker can generate adversarial samples by solving an optimization problem that generates additive noise that maximizes the loss function. Unlike a standard \whitebox attack, in the case of \name the loss function must jointly consider both the \q and the \orig models, as shown in the following equation:
% Our key insight is that this goal can be expressed directly as the objective for the gradient descent. \newtodo{need to polish text} In equation \ref{wb_2}, the first term will be maximized to ensure the confidence score from the \orig model of the correct label is as high  and the second term will be minimized so that the confidence score from the \q model of the correct label is as low.
\vspace{-0.5em}
\begin{equation}\label{eq:whitebox}
    L_{\name}(\theta, x, y) = \model_{orig} \left( x \right)\left[ y \right] -  c \cdot \model_{a} \left( x \right)\left[ y \right]
\end{equation}

\noindent
Here model$_{orig}(x)[y]$ is the raw probability of input $x$ for label $y$ in the \orig model's prediction, and model$_a(x)[y]$ is that for the \q model. Hyperparameter $c$ balances the two probabilities, and is set by default to $c = 1$. $c$ represents a trade off between how well \name evades the \orig model and how well it attacks the \q model. We further study the effect of $c$ in \S \ref{sec:ablation}. Loss function $L_{\name}$ captures the difference between the raw probabilities of the two models. $\theta$ may refer to the parameters of the \orig model, the \q model, or both, depending on how this loss function is used. Plugging this loss function into Equation~\ref{eq:attack} yields the following joint maximization problem:
\vspace{-0.5em}
\begin{equation}\label{eq:dqa}
    A = \argmax_{A \leq \epsilon} L_{\name}(\theta, x + A, y)
\end{equation}
\vspace{-0.2em}
\noindent
that can be solved using stochastic gradient descent or other standard optimization techniques. Our design uses PGD to solve the problem and generate adversarial perturbations. % that mislead the \q model and evade the \orig model.

\begin{figure}[t]
\small
\centering
\begin{subfigure}[t]{0.32\columnwidth}
\includegraphics[width=\textwidth]{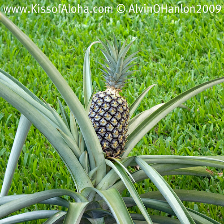}
\caption{Original image.}
\label{fig:original-image-pineapple}
\end{subfigure}~
\begin{subfigure}[t]{0.32\columnwidth}
\includegraphics[width=\textwidth]{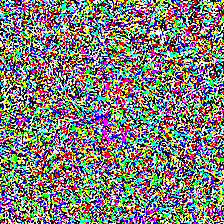}
\caption{Attack noise.}
\label{fig:attack-noise-pineapple}
\end{subfigure}~
\begin{subfigure}[t]{0.32\columnwidth}
\includegraphics[width=\textwidth]{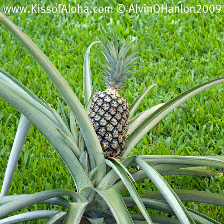}
\caption{Attacked image.}
\label{fig:attacked-image-pineapple}
\end{subfigure}
\caption{Attack against an \q model running on edge device that is undetected by the \orig model. With \name, a pineapple is identified as a cairn by the \q model (confidence 79.2\%), whereas the \orig model still recognizes the image as a pineapple (confidence 76.1\%). The original image is recognized by both the \orig and the \q models as a pineapple (confidence 100.0\% and 98.5\%, respectively).}
\vspace{-1em}
\label{fig:quantized-attack-pineapple}
\end{figure}

\paragraphb{Example.}
We highlight an example of how \name can be used to trick an \q model, depicted in Figure~\ref{fig:quantized-attack-pineapple}. In this example, the adversarial noise (\ref{fig:attack-noise-pineapple}) is added to an image (\ref{fig:original-image-pineapple}), to produce an attacked image (\ref{fig:attacked-image-pineapple}), which is indistinguishable to the human eye from the original image. The attacked image does not affect the prediction of an \orig \resnet50, which accurately predicts the object as a pineapple, but causes \q model (TensorFlow, int8),  running on a resource constrained device, to mispredict the image as a cairn (man-made pile of stones). 

\begin{figure}[t]
\small
\centering
\includegraphics[width=0.8\columnwidth]{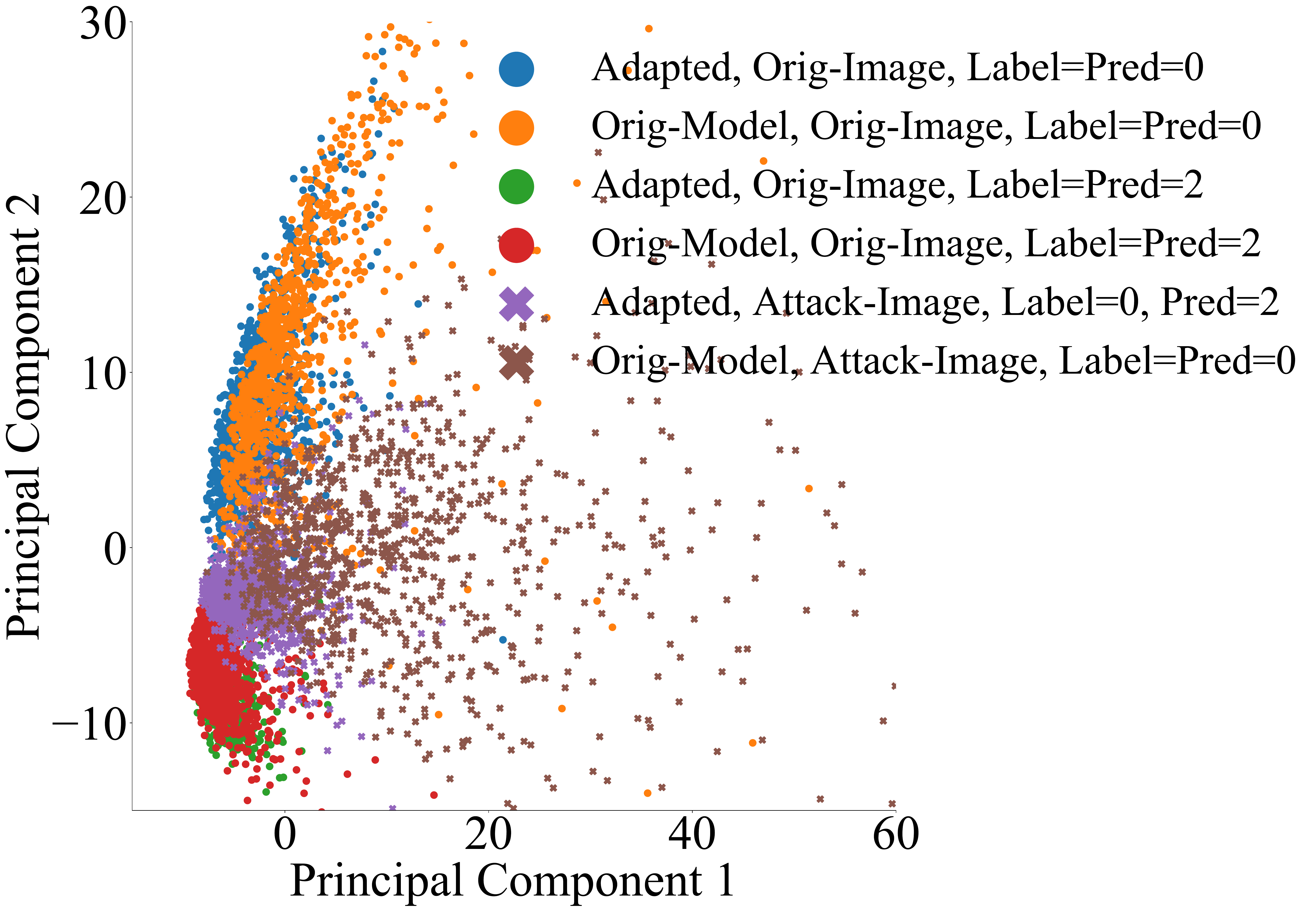}
\caption{PCA visualization of representations learned by the \orig and \q models, using the two most-principal components on MNIST. 
%There are 1,000 original images from the digit 0 and 1,000 images from the digit 2 classes in the MNIST dataset. Both models predict correctly on these images. 
``Adapted, Orig-Image, Label=Pred=0'' indicates the representations learned by the \q model on the digit '0' original images. ``Attack-Image'' shows the representations on the images with \name's adversarial noise.} % (\textbf{Attack-Image}). }
\vspace{-2em}
% for 1000 test samples that belong to either digit ``0'' or ``2'' in the MNIST dataset. ``Orig-Image'' is the unmodified image from MNIST, and ``Attack-Image'' is the image with \name's adverserial noise.
%\newtodo{Asaf: it would be good to: 1. for the purple and brown crosses, include all label 0 predictions, not just 1 in the case of the \q model and 0 in the case of the \orig mode. 2. change the '1' label to '2', sorry about that!. 3. rearrange legend order: first blue, then green, then yellow, then red, then brown, then purple.}} %Class 0 is the correct label for , comparing the  adversarial images of class 0 predicted correctly by both models, which are mistakenly classified to class 1 by \q model but still correctly classified by \orig model after attack. The figure shows representations of 1000 natural (clean) test examples of label 0 from MNIST dataset, 1000 adversarial examples crafted from these images and 1000 natural test examples of label.}
\label{fig:pca-analysis}
\end{figure}

\paragraphb{How does the \name adversarial noise affect the learned representations?}
To better understand how \name's generated adversarial noise affects both the \orig and \q models, Figure~\ref{fig:pca-analysis} presents the Principal Component Analysis (PCA)~\cite{pca} of the representations learned by \resnet50 on 1,000 samples that both the \orig and \q models classify as digit 0 and another 1,000 that both classify as digit 2 from the MNIST hand-written digits dataset~\cite{mnist}\footnote{We use MNIST for PCA visualization because each digit has many samples.}. We pick the activations from the penultimate layer of \resnet50 as the representation of a sample. We similarly analyze the representations after the 1,000 digit 0 images are attacked by \name, which causes the \q model to mispredict digit 2 while the \orig model still predicts correctly.

% \newtodo{1,000 randomly selected samples of the digits 0 and 2 -- Wei/Jiayang: is this correct?} from the MNIST dataset. 

There are several interesting details in the PCA figure. First, even on the original images, there is a subtle difference in the representations learned by the \q and the \orig model, in particular on digit 0 (the blue and orange dots in the upper part of the graph, which belong to the \q and \orig models respectively). Second, the figure shows how \name shifts the representations both for the \q and \orig models. The purple crosses show how \name is able to shift the images belonging to digit 0, from the upper part of the graph to the bottom part. For the \orig model, \name also shifts the points more towards to the bottom of the graph, but less so than it does for the \q model, thus preserving the prediction of the correct label for most of them.

%\newtodo{We investigate the deviations of the model by look at how the natural and adversarial images are represented by different models. We use PCA~\cite{pca} to reduce multidimensional data to lower dimensions for better visualization. 
%PCA Visualization of adversarial images of class 0 predicted correctly by both models, which are mistakenly classified to class 1 by \q model but still correctly classified by \orig model after attack. The figure shows representations of the second to last layer of 1000 natural (clean) test examples of label 0 from MNIST dataset, 1000 adversarial examples crafted from these images and 1000 natural test examples of label.
%Notice that for the natural image of the same class, the clusters of prediction from both models are clearly close and of similar shape. However, for the generated image, 1) the shape of clusters of \orig model and \q model become very different 2) \orig model yields a cluster scattered over the upper and right side of the graph, closer to the clusters of the corresponding natural image, while the cluster of \q model on generated image is concentrated near the clusters of the wrong label. It shows that our attack deviates the predictions of \orig model and \q model differently, in which \q model inclines to extract features that lead to incorrect prediction while \orig model keeps the correct result.}

\subsection{Semi-\blackbox \name Attack}
\label{sec:semi-blackbox}

In the semi-\blackbox attack, the attacker has access to the \q model but not the \orig model. We assume it is substantially easier for the attacker to obtain one of the edge devices with the \q model than hacking into a server with the \orig model. Practically, an attacker can obtain the adapted model from an edge device and recover the differentiable quantization model by extracting the zero points, scales and weights for each layer in the downloaded model, and retain its accuracy without any fine-tuning. However, without access to the \orig model parameters and training data, the gradient calculations in \whitebox attack no longer apply.

\begin{figure}[htp!]
\centering
\includegraphics[width=0.35\textwidth]{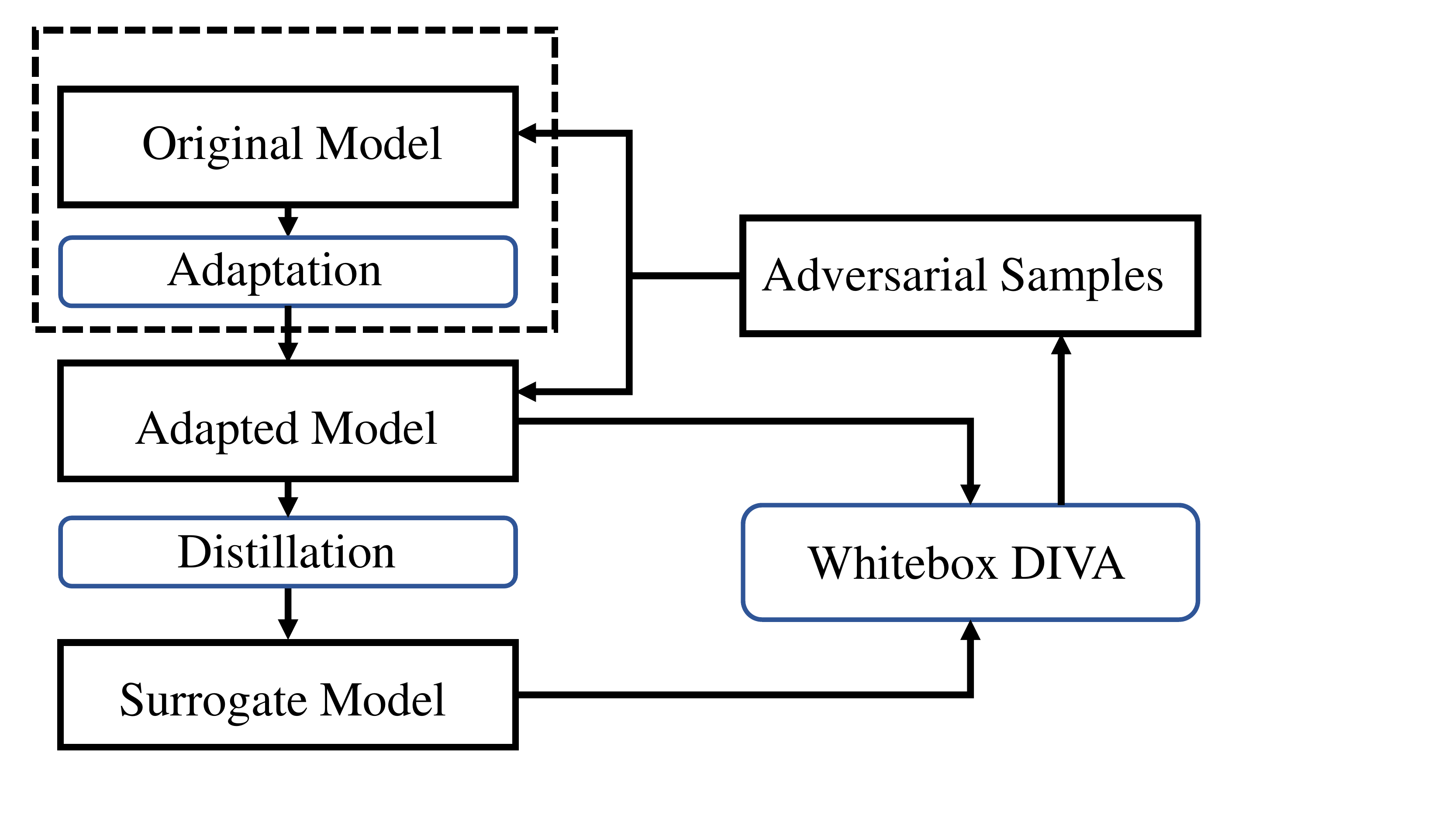}
\vspace{-1em}
\caption{\name semi-\blackbox attack pipeline. The dotted box indicates components that the attacker has no access to.}%Once the attacker gains access to the \q model, they run distillation to train a surrogate \orig model, and launch whitebox \name against these two models. The resultant adversarial samples can be used to attack the original \orig and the \q models.}
\vspace{-3em}
\label{semi_bb_pipeline}
\end{figure}

Fortunately, attacks are frequently transferable~\cite{papernot2016transferability}: adversarial samples generated for one model often transfer to another model for a similar task. To launch a semi-\blackbox attack, we reconstruct a \fp surrogate model that behaves similarly as the \orig model and apply $L_{\name}$ to generate adversarial samples for the \q and the surrogate model. Figure~\ref{semi_bb_pipeline} shows the semi-\blackbox attack pipeline.

We reconstruct a surrogate model as follows. Based on the \q model, we create an architecture that matches the \q model. %except with quantization-aware operators removed.
The surrogate model's parameters are initialized using the pretrained ImageNet parameters from TensorFlow\cite{keras-applications} when possible or the parameters of the \q model.% except with full precision. 
We then \finetune the surrogate model via knowledge distillation. Unlike typical knowledge distillation that trains a model with less precision using an \orig model, we use knowledge distillation to create semi-\blackbox attack. In our case, \name treats the \q model as the teacher and the surrogate  model as the student. We train the surrogate model to minimize the loss between its prediction and the label predicated by the \q model while minimizing the distillation loss~\cite{hinton2015distilling}.

\vspace{-1em}
% After training this distilled \orig model, we use this model to generate the attack, defined in equation \ref{semibb},  whose targets are the original \orig model and the quantified model. 

\subsection{Blackbox \name Attack}
\label{sec:blackbox}
In the full blackbox setting, the attacker does not have access to neither the entire \fp model nor the parameters of the \q model. A fully-blackbox attack can be generated by reconstructing both a surrogate \fp model, described in \S\ref{sec:semi-blackbox}, and a surrogate \q model followed by adaptation and fine-tuning. The \blackbox attacks then can be generated using the surrogate models and evaluated on the \orig \fp and \orig \q model.

\section{Evaluation} \label{sec:eval}

This section presents an evaluation of \name. It first describes the experimental setup (\S\ref{sec:experimentalsetup}), then the efficacy of the \name \whitebox, semi-\blackbox and \blackbox attacks (\S\ref{sec:eval-whitebox}), and of a defense method (\S\ref{sec:eval-defense}), and finally show that \name generalizes to pruning (\S\ref{sec:pruning}). 
% and finally the efficacy of \name against various defense methods (\S\ref{sec:eval-defense}).

\subsection{Experimental Setup}
\label{sec:experimentalsetup}

\paragraphb{Datasets.}
Our main dataset contains 50,000 images from the ImageNet Object Localization Challenge of 2012-2017~\cite{imagenet-localization}. We randomly select 40\% of the 50,000, or 20,000, images from the dataset as our training dataset. For the validation datasets, we randomly select 3,000 images from the remaining 30,000 images covering all 1,000 classes of ImageNet, with an average of three images perclass. When selecting these 3,000 validation images, we ensure that they are correctly classified by all relevant models and architectures. Eventually we created three sets of 3,000 validation images: one dataset for quantized whitebox/semi-\blackbox attacks, one for quantized \blackbox attack, and one for the attack on pruned models.

The semi-\blackbox and \blackbox attacks also require additional data to train the surrogate models. Therefore, we use 1\% (12,811) images from the training dataset of the same ImageNet challenge to train the surrogate models. Since the semi-\blackbox and \blackbox attacks assume that the attacker cannot access any training data of the \orig model, we ensure that the 12,811 images have no overlap with our main dataset by selecting the 50,000 images in our main dataset from the validation dataset of the ImageNet challenge.

% We pick the first 40\% of the 50,000 images from the dataset to \finetune our models. For the validation dataset, we selected 3000 images randomly from the remaining 60\% the 50,000 images, where all 9 of our evaluation models in the attack evaluation section (original model, \q model and semi-\blackbox reconstructed model) were able to classify the image correctly. The validation dataset covers all 1,000 labels of ImageNet, with an average of three images per class. In addition, the distilled \orig model in the semi-black box setting is trained separately on the remaining 12,811 images, \newtodo{why this number: 12,811?}.

\begin{figure*}[t!]
\small
\centering
\begin{subfigure}{0.25\textwidth}
\includegraphics[width=\textwidth]{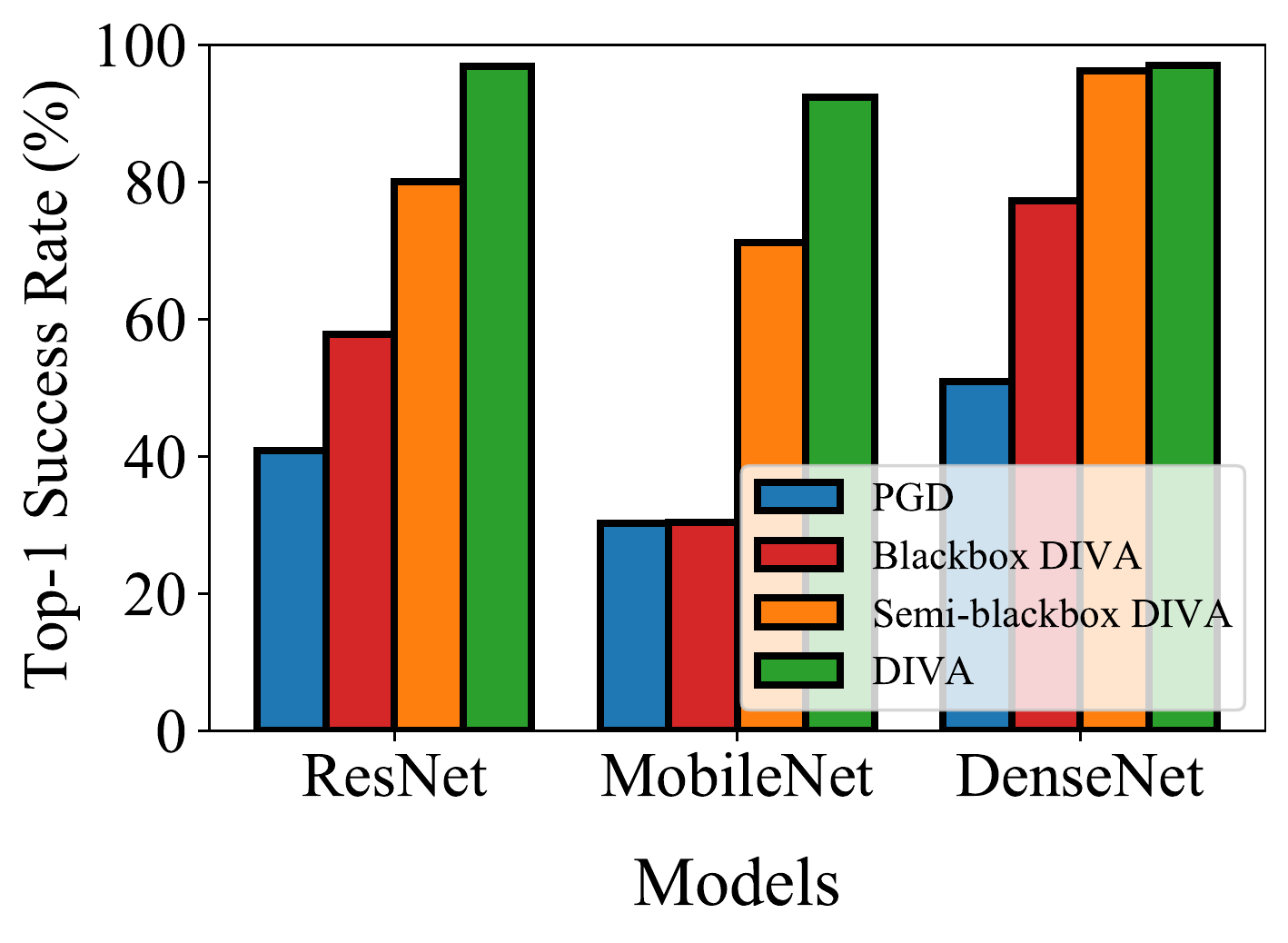}
\caption{Top-1 success rate.}
\label{fig:top-1-attack}
\end{subfigure}~
\begin{subfigure}{0.25\textwidth}
\includegraphics[width=\textwidth]{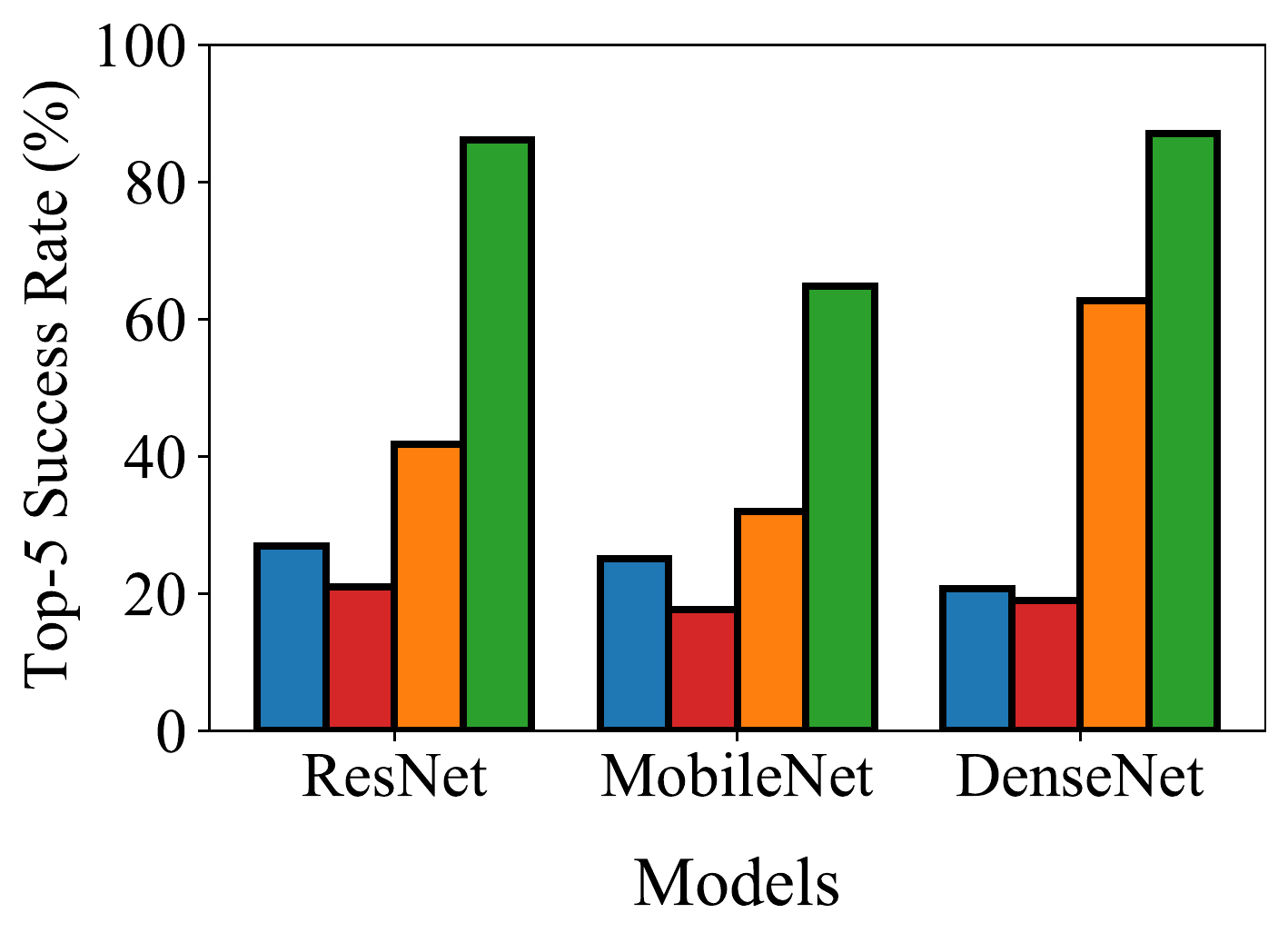}
\caption{Top-5 success rate.}
\label{fig:top-5-attack}
\end{subfigure}~
\begin{subfigure}{0.25\textwidth}
\includegraphics[width=\textwidth]{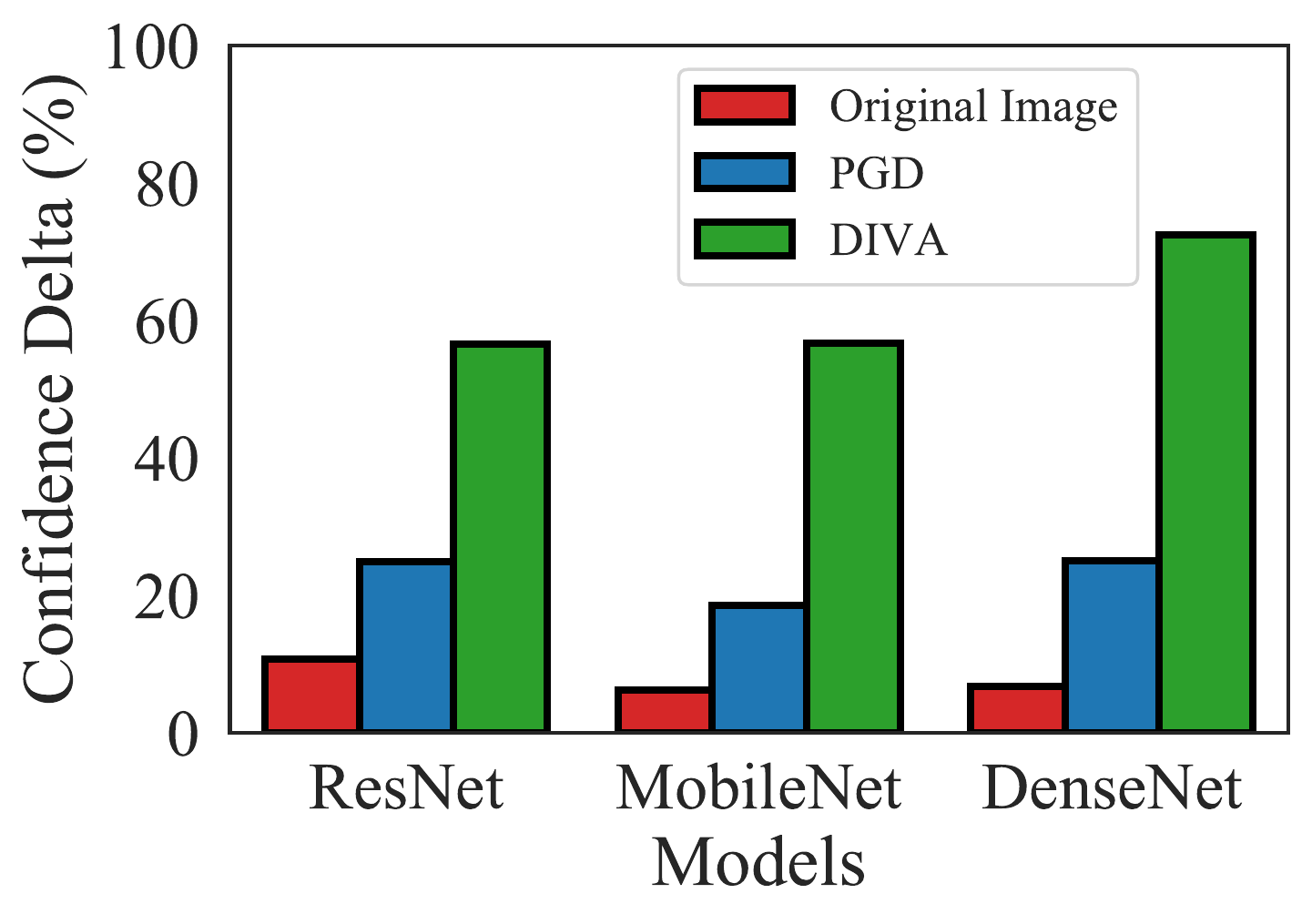}
\caption{Confidence delta for top-1.}
\label{fig:confidence-delta}
\end{subfigure}~
\begin{subfigure}{0.25\textwidth}
\vspace{+0.2em}
\includegraphics[width=\textwidth]{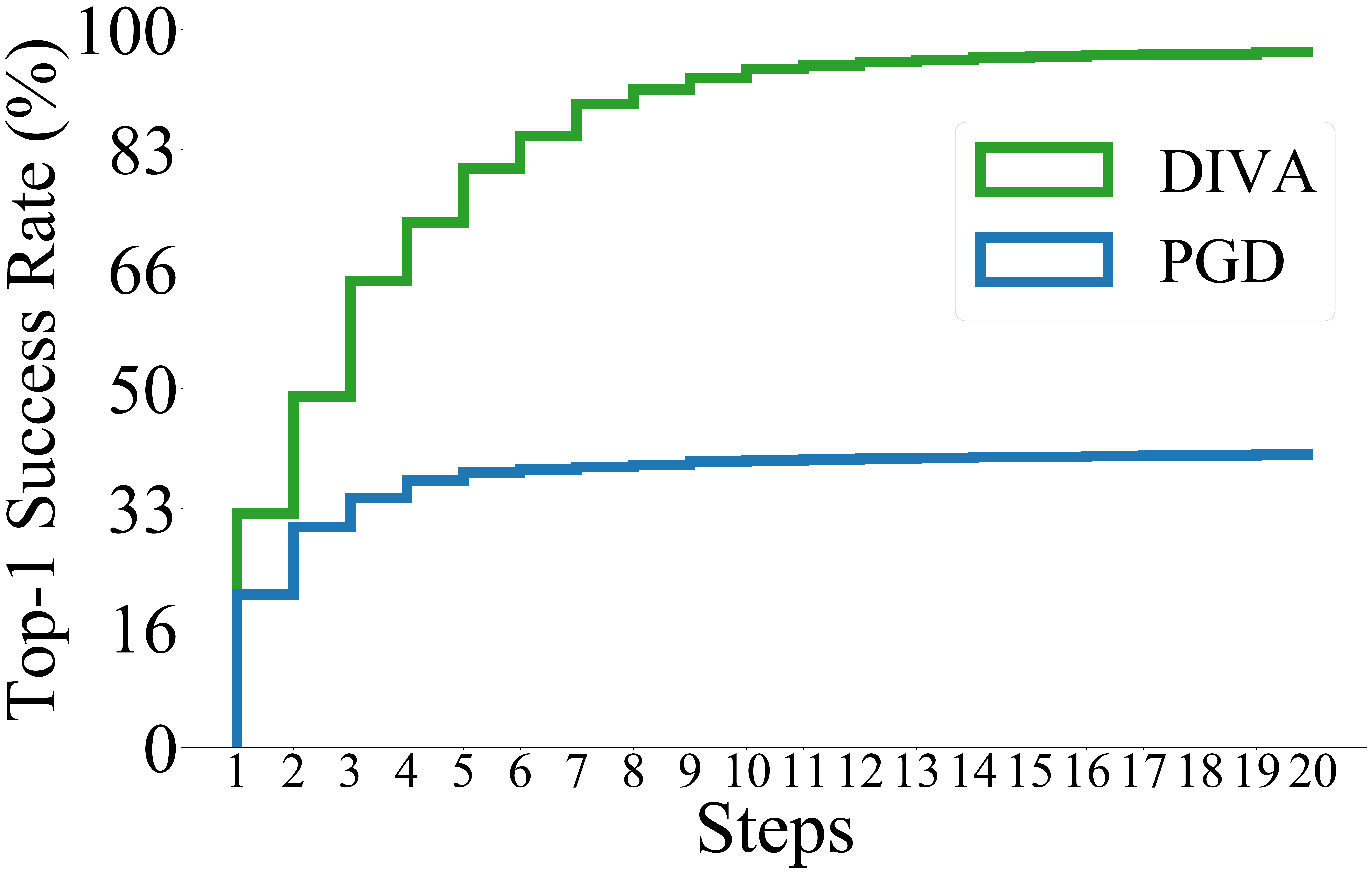}
\vspace{-0.3em}
\caption{Top-1, varying attack steps.}
\label{fig:step_vs_successrate}
\end{subfigure}
\vspace{-1em}
\caption{\textbf{Attacks on quantized models.} }%\name compared to PGD, with three different models under the top-1, top-5 and confidence delta metrics, and as a function of.} %In both \whitebox and semi-\blackbox scenarios, \name significantly outperforms PGD in the top-1 and top-5 metrics. Under the \whitebox attack setting, \name significantly reduces the average confidence delta in the correct label, compared to PGD.}
\vspace{-1em}
\label{fig:top-n-comparison}
\end{figure*}

\paragraphb{Models.}
We use three architectures in the evaluation: \resnet50, \mobilenet and \densenet121. For each one, we create four models: the \orig model, the \q model, the surrogate \orig model for the semi-\blackbox attack and the surrogate \q model for the \blackbox attack.
We create the \orig models by downloading pre-trained models from TensorFlow Keras Applications~\cite{keras-applications} and finetuning them on our training dataset of 20,000 images. We use pre-trained models instead of training from scratch to save resources and ensure that the models evaluated have state-of-the-art accuracy. The finetuning step is necessary for model adaptation, in the case when the dataset used for finetuning is a subset of the training dataset, not a disjoint dataset. We validate that the generated \orig models achieve the state-of-the-art accuracy.

We generate quantized models by first applying TensorFlow Model Optimization \texttt{tfmot}'s \texttt{quantize\_model}~\cite{quantize-model} on the \orig models using int8 quantization. We then apply QAT to these models on our training dataset of 20,000 images. After two epochs of QAT, the quantized models reach validation accuracy comparable to their corresponding \orig models. We observe that more epochs do not improve accuracy but worsen the stability between the \orig and the \q models. %, so we stop QAT after two epochs. 
The surrogate models are created by applying knowledge distillation and QAT with the aforementioned 12,811 images.

We generate two types of pruned models on the three architectures: (1) by applying Keras weight pruning on \orig models, and (2) by taking the pruned models and then quantizing them using the same pipeline we used for quantized models, while preserving the sparsity using \texttt{tfmot}. Both types of models are then fine-tuned to reach their highest accuracy on the dataset. After pruning, the model sizes were compressed to one third of their original size.

% The semi-\blackbox model is constructed using a distillation method in which we train the model by including a distillation loss, in addition to its regular loss, which is calculated using the predictions of the \q model.

% For all three neural network architectures, we import the pre-trained weights, which were trained on ImageNet using TensorFlow, and \finetune them for two epochs on the training dataset of 20,000 images. We then apply QAT on these \orig models on the same dataset. We apply QAT for two epochs so that the \q model reaches a validation accuracy comparable to the \orig model. Further, we noticed that training for more epochs will lead to no to little accuracy improvement but a relatively higher drop in the model's stability. For the reconstructed \orig models in the semi-\blackbox setting, we apply knowledge distillation where the \q models act as the teachers and the constructed models are the students. We train this distillator on the disjoint set of 12,811 images so that the distilled model does not see the dataset where the original \orig model and the \q model are trained on. This is because we assume that both the \orig model and the dataset it is trained on should be inaccessible to the hacker in the semi-\blackbox setting.

\paragraphb{Attack construction.}
In our experiments, we construct the PGD attack targeting the \q model as the baseline attack. We choose $\epsilon = 8$ for both baseline and \name attack, as it has been shown that the perturbation of $8/255$ is generally imperceptible to human eyes. We run both the baseline and the \name attacks with a step size $\alpha = 1$~\cite{defensive_q}, and set the maximum number of steps $t = 20$. We do not initialize the attack using random noise because random start is less effective in a single run. Instead, we initialize with a natural sample.

% \paragraphb{Defense construction.}

% For all Minimax robust training defenses, we use the RMSprop optimizer \cite{hinton2012neural}, which is used by default in TensorFlow. We configure RMSprop with learning rate $2\times 10^{-5}$ and discounting factor $\gamma = 0.9$, following the best reported hyperparameter settings~\cite{Ruder16}. We set the batch size to $16$, the maximum that fits our GPU's memory. For Minimax PGD Robust Training, we train the \orig model for 10 epochs until validation loss converges. We then quantize the model and apply QAT using the clean training dataset for two epochs for fair comparison with the undefended \q model. For Minimax DQA Quantization  Aware Training, we train the \q model also for 10 epochs until the validation loss converges. For distillation, we use the \orig model as the teacher and the \q model as the student. We set the distillation temperature in the softmax to $10$ and assign the student model loss with a weight of $0.1$ (and the distillation loss with a weight of $0.9$).

\paragraphb{Success metrics.}
We define a successful attack as one where both: (a) the attack did not cause the \orig model to mispredict, \ie if it correctly classified the original image, it correctly classifies the perturbed one; and (b) the attack caused the \q model to mispredict, \ie the \q model correctly classified the natural image, but incorrectly classified the perturbed one.
We define three metrics to quantify \name's efficacy. The first, which we refer to as \emph{top-1 success rate}, measures whether the attack caused the top-1 prediction of the \q model to be incorrect. The second, \emph{top-5 success rate}, measures whether the attack caused the \q model's top-1 prediction not to even appear in the top-5 predicted classes of the \orig model.

We define a third metric, \emph{confidence delta}, which measures the difference between the confidence of the correct class between the \orig model and the \q model, on the attacked image. For example, if the correct class in an image is a pineapple, and the confidence of the \orig model on the attacked image is 80\%, while the confidence of the \q model on the attacked image is 20\%, the confidence delta will be equal to 60\%. %Note that the confidence delta can be negative, in the case when the \q model has a higher confidence in the correct class than the original \orig model. We measure the average confidence delta between the \orig and \q models across a large number of images. 
We additionally measure DSSIM~\cite{hore2010image}, which quantifies the similarity between two pictures from human's perspective, to examine whether the adversarial samples are similar to their corresponding natural samples.

\paragraphb{Machine.} Experiments are conducted on a server with 4 Intel 20-core Xeon 6230 CPUs, 376 GB of RAM, and 8 Nvidia GeForce RTX 2080 Ti GPUs with 11 GB of memory.

%\paragraphb{Ethical considerations.}
%This study only focuses on public datasets and models, and does not involve either sensitive or private data.

\subsection{\name with Quantization}
\label{sec:eval-whitebox}

%In our experiments, we construct the pgd attack which maximize the cross-entropy loss between the correct prediction and the prediction from the \q model as our baseline method as it is one of the strongest attack method. We chose $\epsilon = 8$ as it has shown that the perturbation of 8/225 is imperceptible to human eyes. We run the PGD attack with $step  size = 1$ following \cite{defensive_q} and run the max number of $steps = 20$ or until the attack succeeds, meaning when the attack cause the \q model miss-predict but is undetectable to the \orig model.

\paragraphb{DSSIM.}
We measure DSSIM among all generated adversarial images on three model architectures comparing the PGD baseline against \name. The resulting DSSIM for all images are below 0.0092, which means that both the PGD and \name attacks are imperceptible to humans.

\paragraphb{Top-1 and Top-5.}
Figure~\ref{fig:top-n-comparison} compares the performance of \name with PGD, under the three architectures. Figure~\ref{fig:top-1-attack} shows that \name significantly outperforms the baseline attack method, PGD, both for the \whitebox and semi-\blackbox scenarios. The \whitebox attack is able to cause the \q model to misclassify the correct label, while maintaining the \orig model's correct classification, for 92.3--97\% of the validation dataset. PGD is also able to successfully attack some of images, because in general the \q model is less robust to noise than the \orig model. However, its attack success rate is much lower than \name: it achieves a 30.2--50.9\% top-1 success rate. As illustrated previously in Figure~\ref{fig:pgd_vs_dqa}, while the PGD baseline is able to cause the \q model to mispredict, the adversarial samples that it generates also cause the \orig model to mispredict, a key reason that it is substantially less effective than \name. 

% further shows the distributions of images after both attacks on \resnet50. Since \name is focusing on only making the \q model miss-predict while keeping the \orig model correct, it either succeeds or it fails in the case where the \q model and the \orig model predict both correct or both wrong. On the other hand, the baseline attack is able to attack the \q model for most of the images. But as a side effect, it also attacked the full-presicion model and the portion of these attacks is very high. This proves our theory of why PGD is not suitable for our attack purpose.

Figure~\ref{fig:step_vs_successrate} plots the top-1 success rates for PGD and \name on \resnet50 as the number of attack steps increases. \name outperforms PGD starting from step 1. PGD plateaus at 40.8\% on step 7, whereas \name reaches 96.9\% on step 11.
For the top-5 success rate, the differences are even more stark, as \name in the \whitebox setting is able to achieve a 2.6-4.2$\times$ higher top-5 success rate than PGD.
As expected, the semi-\blackbox attack is less successful than the \whitebox one, since \name does not have access to the original \orig model. However, it is still able to outperform PGD across all the experiments. It has a high success rate especially for the top-1 metric, successfully attacking 71.1-96.2\% of the images across the three models.

Blackbox \name does not perform as well as the semi-\blackbox version, resulting in a top-1 success rate of 30.3--77.2\%, which is higher than PGD's. However the blackbox attack's top-5 success rate is 17.6--21\%, which is lower than PGD's. Since the blackbox attack has to generate an extra surrogate model compared to the semi-blackbox version, it is harder for it to locate the decision boundaries. We further noticed that the top-1 success rate for \blackbox \name on MobileNet is worse than the other two networks, when comparing the improvements with their PGD baselines. This is because small under-parameterized neural networks often generalize worse than larger networks~\cite{novak2018sensitivity}, leading to poor transferabilty of \name from the surrogate models to \orig models.

\paragraphb{Confidence delta.}
Figure~\ref{fig:confidence-delta} presents the average shift in confidence for the correct label under the three models, comparing \name (in a \whitebox setting) to PGD. For each model, we first compare the average confidence delta of the original image, between the \orig model and the \q one. As expected, in all three models, there is a relatively modest difference in the confidence delta solely due to quantization (on average only 7.9\%).
As expected, the attacks cause the confidence delta to shift more. For PGD, the average confidence delta varies between 18.6--25\% across the three models. For \name, the average confidence delta varies between 56.6--72.4\%. The much higher shift in confidence for the correct prediction explains why \name has a much higher top-1 and top-5 success rate than PGD.

\paragraphb{Attack speed.} We measure the wall-clock time for both PGD and \name with our experimental setup. They run at almost the same speed of one second per step.

\begin{table}[t]
\centering
\small
\begin{tabular}{|l|r|r|}
\hline
& \multicolumn{1}{|l|}{PGD} & \multicolumn{1}{|l|}{\name}  \\
Architecture & \multicolumn{1}{|l|}{Attacking Quantized} & \multicolumn{1}{|l|}{Attacking Quantized}  \\
\hline
\hline
\resnet50 & 98.7\% & 97.0\%\\
\hline 
\mobilenet & 98.7\% & 95.1\%\\
\hline
\densenet121 & 98.4\% & 96.7\%\\
\hline
\end{tabular}
\caption{Comparing attack success rate solely against \q models. For pruning, \name and PGD both achieve 100\%. For pruning + quantization, PGD has success rates of 98.4--99.7\% and \name 98--99.7\%.}
\label{tab:successrateoquantized}
\end{table}

\begin{figure}[t!]
\centering
\includegraphics[width=0.48\textwidth]{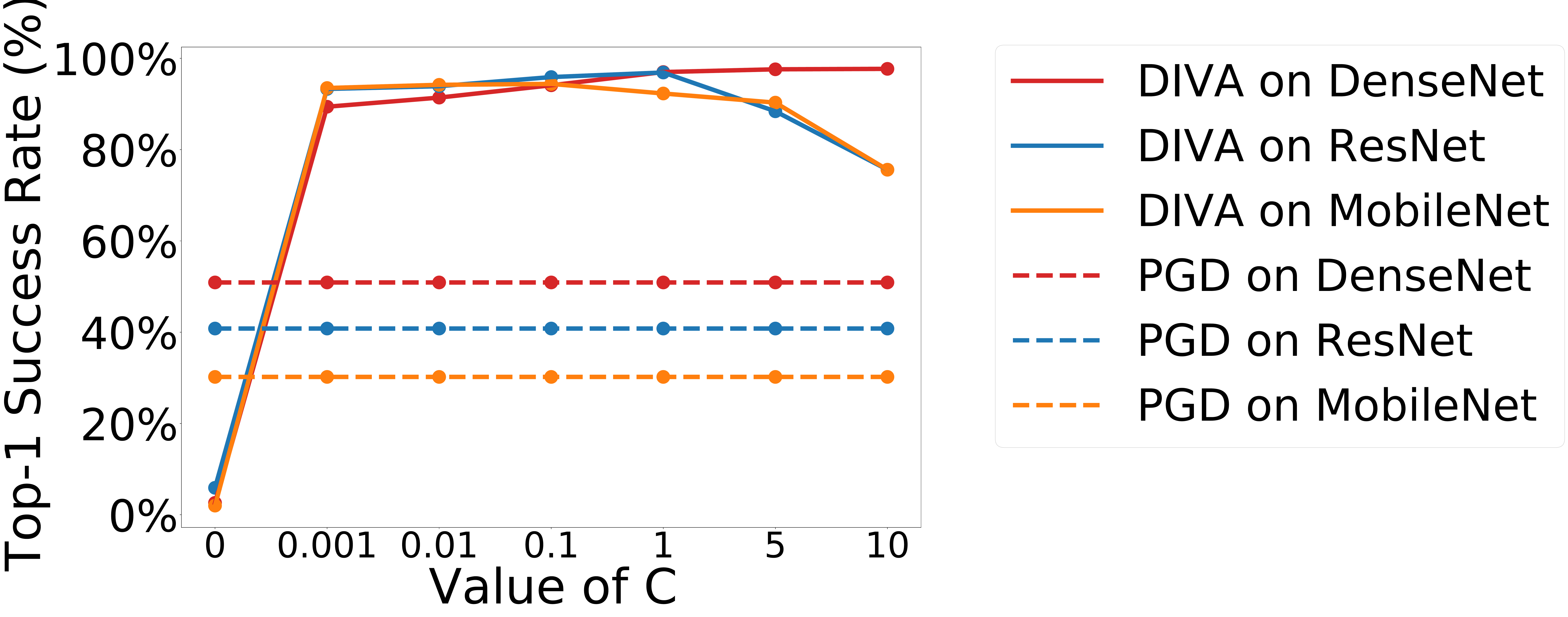}
\vspace{-2em}
\caption{Whitebox \name with varying $c$.}
\label{fig:ablation_study}
\vspace{-1.5em}
\end{figure}

\paragraphb{Evasion cost.} We measure the cost of evading the \orig model, by comparing the success of \name at attacking the \q model compared to PGD. Note that we generate the adversarial samples using \name as usual, considering both the \orig and \q models. However, to compute the success rate, the only criterion is that an adversarial sample misleads the \q model.
The results are presented in Table~\ref{tab:successrateoquantized}, and show that, despite the constraint not to affect the \orig models, \name achieves a very high attack success rate on the quantized models, but falls slightly short of PGD (1.7--3.6\% less effective). When attacking the pruned models, which we describe in more detail in \S\ref{sec:pruning}, \name is almost as effective as PGD and just falls negligibly short of PGD in few cases (0.2--0.4\% less effective). %\newtodo{delete: The evasion cost can be further reduced by tuning the $c$ parameter (Equation~\ref{eq:whitebox}), at the expense of a lower evasion success rate.}

\begin{figure*}[t!]
\small
\centering
\begin{subfigure}{0.24\textwidth}
\includegraphics[width=\textwidth]{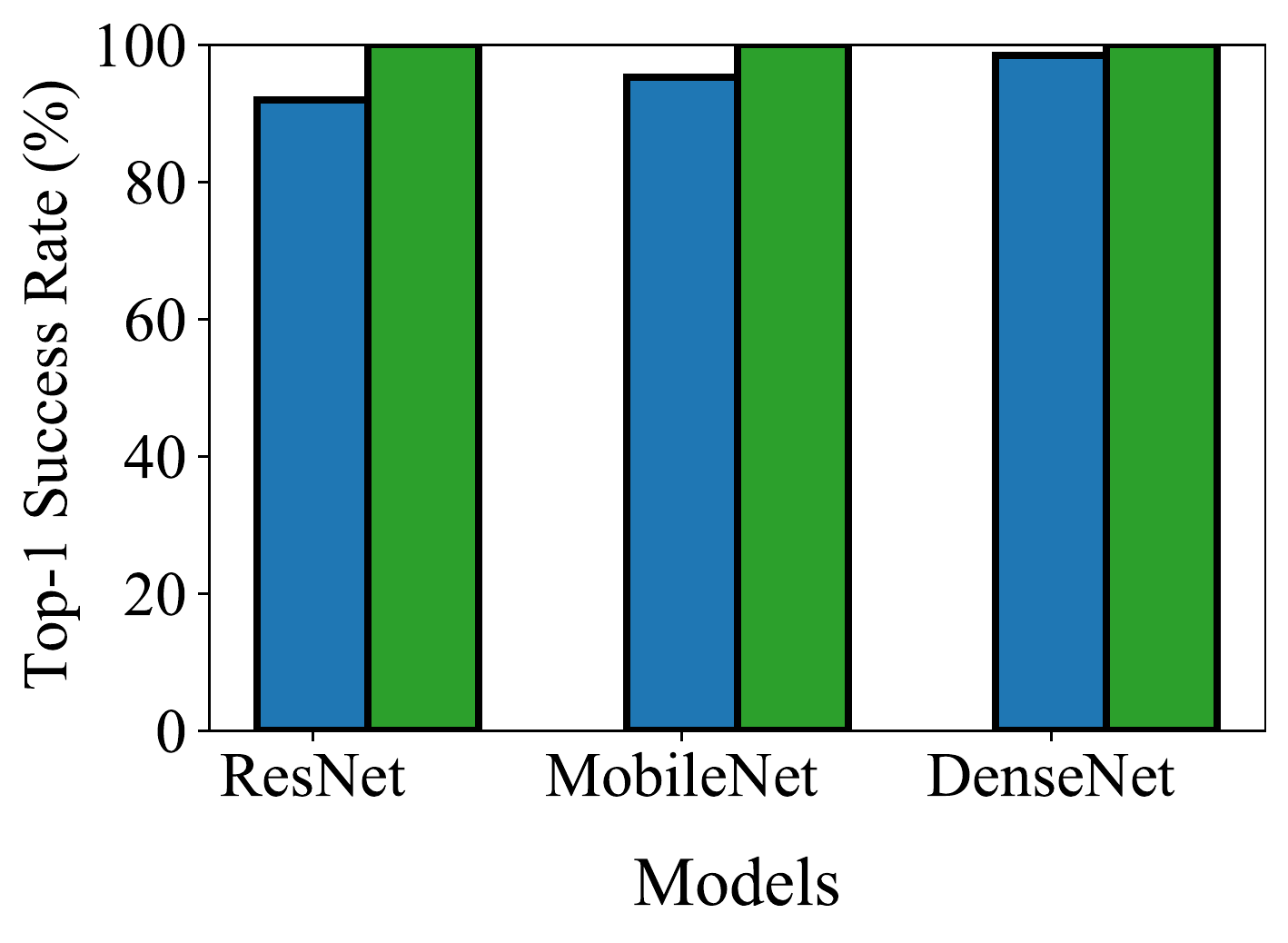}
\caption{Top-1 success rate.}
\label{fig:top-1-attack-pruned}
\end{subfigure}~
\begin{subfigure}{0.24\textwidth}
\includegraphics[width=\textwidth]{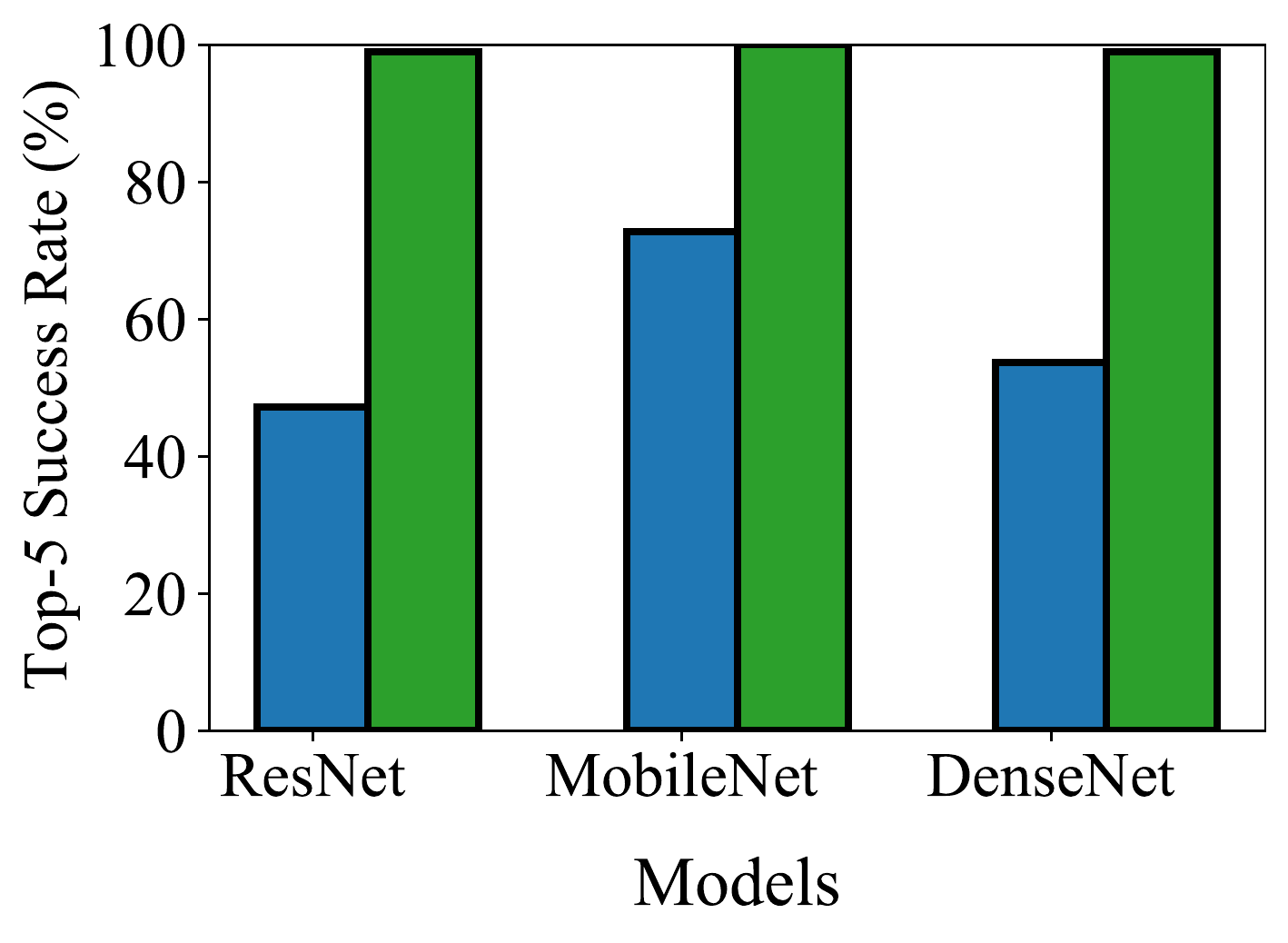}
\caption{Top-5 success rate.}
\label{fig:top-5-attack-pruned}
\end{subfigure}~
\begin{subfigure}{0.24\textwidth}
\includegraphics[width=\textwidth]{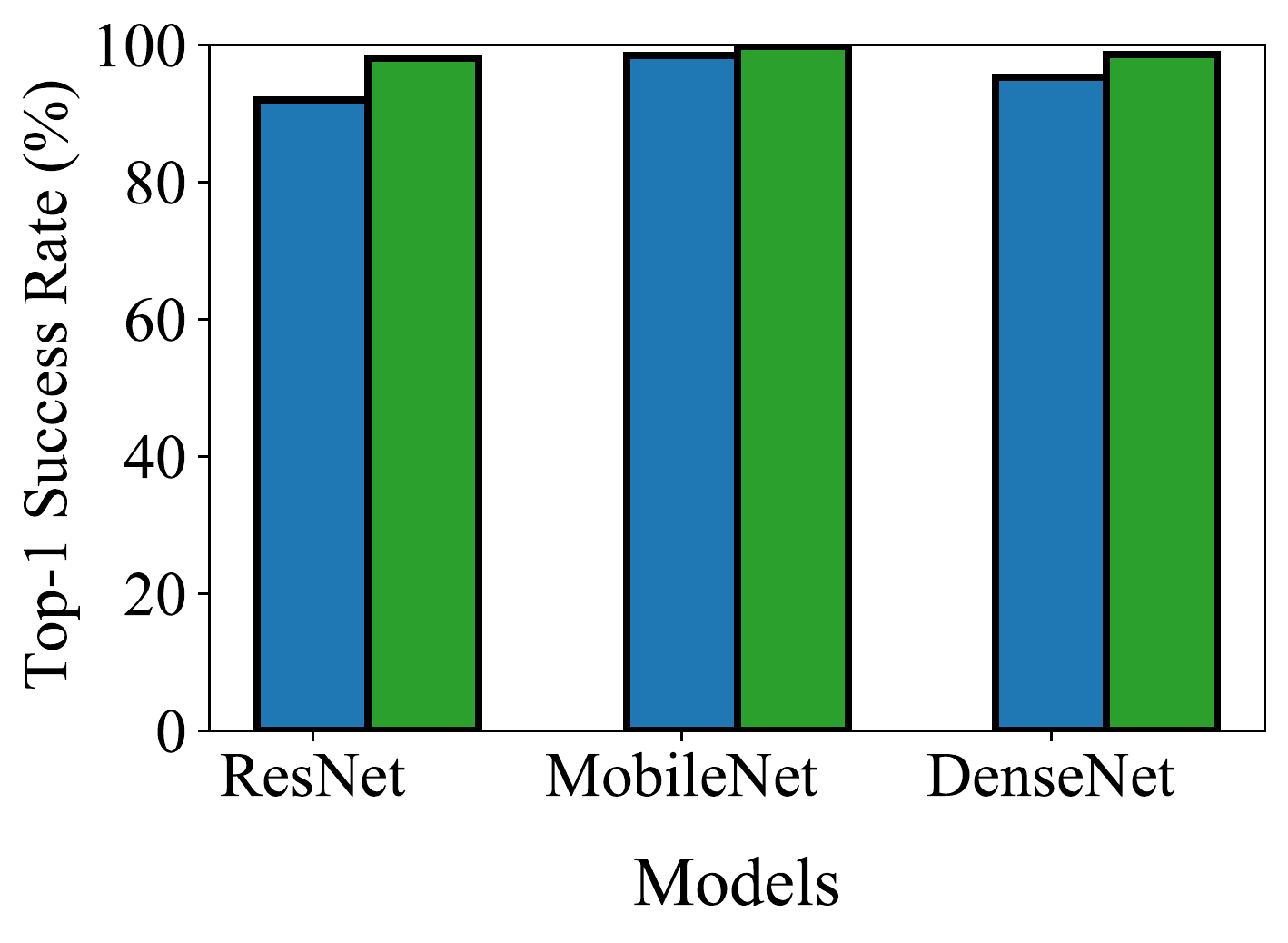}
\caption{Top-1 success rate.}
\label{fig:top-1-attack-pruned-qat}
\end{subfigure}
\begin{subfigure}{0.24\textwidth}
\includegraphics[width=\textwidth]{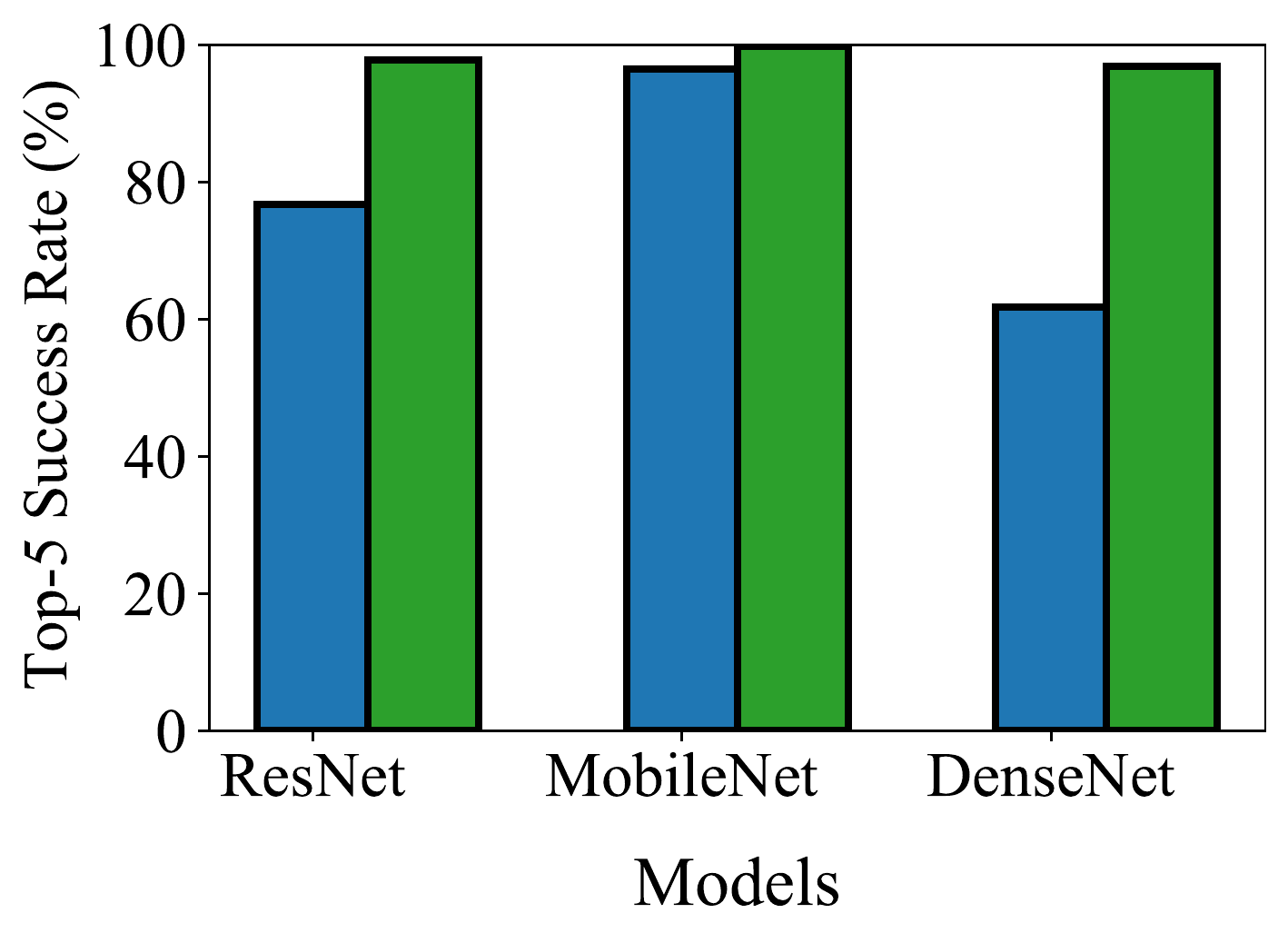}
\caption{Top-5 success rate.}
\label{fig:top-5-attack-pruned-qat}
\end{subfigure}
\vspace{-0.5em}
\caption{\textbf{Attacks on pruned models.} %\name compared to PGD, with three different models under the top-1 and top-5 metrics. 
(a,b) Pruned models. (c,d) Pruned and quantized models. }
\vspace{-1em}
\label{fig:top-n-comparison-pruning}
\end{figure*}

\subsection{Balancing Between Evading and Attacking}\label{sec:ablation}
The hyper-parameter $c$ is used to balance the effect of the two loss terms in Equation \ref{eq:whitebox}. A small $c$ focuses the attack on not being detected by the original model, and a large $c$ focuses on attacking the adapted model. %Since adversarial attacks are transferable, a $c$ value that is set to be too small will fail to attack the quantized model and if it is too large it will expose the attack to the original model.

Under the setting of quantization, we conduct an ablation experiment for $c=\{0,0.001, 0.1,1,5,10\}$ with whitebox \name . Figure \ref{fig:ablation_study} shows the top-1 attack success rate reaches a peak at $\{96.9\%, 94.4\%, 97.7\%\}$ when $c= \{10, 1, 0.1\}$, respectively for each network we study. The results show that the attack achieves a relatively high success rate for the studied models for $c=[0.001,1]$. We set $c = 1$ by default because it provides the highest average top-1 success across the three model architectures.

The success rate on solely attacking the adapted model is increased to \{97.6\%, 98\%, 97.7\%\}, respectfully for each model architecture when $c = 10$, compared to the numbers in Table  \ref{tab:successrateoquantized} when $c = 1$. The results show that the evasion cost can be reduced by tuning $c$ at the expense of a lower average evasion success rate.

\subsection{Other Baseline Attacks}
\label{sec:otherbaselines}
We also evaluated other baseline methods: the $L_{\infty}$ CW attack \cite{carlini2017towards} and Momentum PGD \cite{PDGm} attack, under the quantization setting and the top-1 success criterion. We follow the same hyperparameter setup as the CW attack in Madry et al~\cite{PGD}. For Momentum PGD, we use a momentum term of 0.5 in addition to the standard PGD attack as this gives the best performance.
Our results show that across the three architectures, CW achieved an average success rate of 25.5\% and Momentum PGD achieved 39.4\%, both of which are worse than PGD (40.6\%).

\subsection{Robust Training as a Defense }\label{sec:eval-defense}

As explained in \S\ref{sec:robust}, robust training can effectively reduce the success rates of adversarial attacks such as PGD. This subsection compares the effectiveness of the PGD minimax robust training on both PGD and \name. %For the comparison to make sense, we count an \name attack as successful when the generated adversarial sample misleads the \q model despite that \name needs to consider both \q and \orig models when generating the sample.%

Since the original PGD robust training code~\cite{PGD,robust-library} is implemented in Pytorch, we ported \name to Pytorch for fair comparison. We used the default hyperparameters used by the authors~\cite{robust-library} ($\epsilon = 8/255 \approx 0.03$, an attack learning rate of 0.00375 with 20 attack steps without random start).
% We measure \name and PGD's adversarial accuracy against quantized models, after applying PGD robust training to test whether \name remains effective after robust training. The experiment is conducted using a standard Pytorch robust training library~\cite{robust-library} with its default attack parameters ($\epsilon = 8/255 \approx 0.03$, an attack learning rate of 0.00375 with 20 attack steps without random start). 
%for convenience as the robust training library we used \cite{https://github.com/MadryLab/robustness} is in pytorch. 
We used the pre-trained robust Resnet50 model in the library as the \orig model and generated the corresponding \q model using the PyTorch-Quantization toolkit~\cite{pytorch-quantization}. 

Our results show that when we use $c = 5$ in \name, it improves the top-1 success rate from 10.5\% to 12.8\%, compared to PGD. Meanwhile, the evasion cost of \name is as similar as PGD, or even slightly better, resulting a robust accuracy of 22.63\% for PGD, and 21.77\% for \name on quantized model. We found the best $c$ value for generating the evasive attack under robust training is $c=1.5$. This value produces a 4\% lower success rate on attacking the adapted model, but the overall evasive attack success rate is 10.1\% higher compared to PGD. Both the evasive attacks' success rates for PGD and DIVA drop when they attack the robust trained models. We think this is because the non-overlapping area between the decision boundaries of the adapted model and the original model becomes smaller, due to the fact that they are both trained to cover the worst case attacks.

\subsection{\name against Pruning Adaptation}
\label{sec:pruning}
 
We evaluate whitebox \name on pruned models. Figure~\ref{fig:top-n-comparison-pruning} shows that under both the top-1 and top-5 metrics, \name achieves a success rate of 97.8\% or higher in all cases and always performs better than PGD. The top-1 success rates of PGD and \name are much closer than they are in the quantization setting potentially because pruning makes more intrusive changes to the model weights than quantization. The larger gap between the pruned and the \orig models, exhibited by the high instability ranging from 17.1--33.5\% and the high confidence delta on the original images ranging from 10--36.1\%, allows PGD to attack the pruned models without collaterally damaging the \orig models. However, \name increases the confidence delta on the perturbed images by 8.3--16\% more than PGD. The top-5 success rates of \name in most settings are significantly higher than PGD.
\section{Case Study: Face Recognition}
\label{sec:casestudy}

\begin{figure}[t]
\small
\centering
\begin{subfigure}[t]{0.32\columnwidth}
\includegraphics[width=\textwidth]{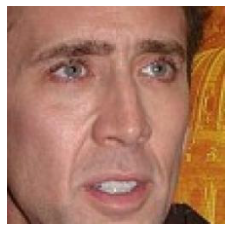}
\caption{Original image.}
\label{fig:original-image}
\end{subfigure}~
\begin{subfigure}[t]{0.32\columnwidth}
\includegraphics[width=\textwidth]{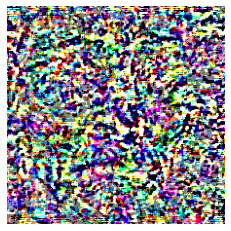}
\caption{Attack noise.}
\label{fig:attack-noise}
\end{subfigure}~
\begin{subfigure}[t]{0.32\columnwidth}
\includegraphics[width=\textwidth]{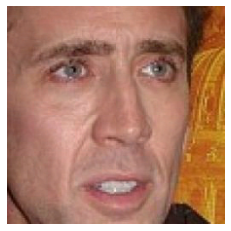}
\caption{Attacked image.}
\label{fig:attacked-image}
\end{subfigure}
\vspace{-1em}
\caption{\name causes the \q model to misidentify Nicolas Cage as Jerry Seinfeld with 95.7\% confidence, whereas the \fp model still recognizes the face as Nicolas Cage with 93.4\% confidence. The original face is recognized by both models as Nicolas Cage with 100\% confidence.}

\label{fig:targeted-attack}
\end{figure}

We present the results of our case study that uses \name to attack a face recognition model, whose quantized version represents a model that would be running on an edge device (\eg security camera or phone). %In the case study we not only cause the model to mispredict, but cause it to predict a person of our choice. 
For example, \name causes the model to misidentify Nicolas Cage as Jerry Seinfeld as shown in Figure~\ref{fig:targeted-attack}.

\paragraphb{Dataset.}
Our dataset includes 11,640 images belonging to 150 people from the PubFig database~\cite{Pubfig}. We pick the first 90\% of the images from the dataset to \finetune our models. For the validation dataset, we randomly select 450 images from the remaining 10\% of the dataset that both the \fp and the \q models classify correctly. The validation dataset covers all 150 labels of our dataset, with three images per class.

\paragraphb{Models.}
We use the VGGFace model, which internally employs \resnet50's architecture, and evaluate it in the \whitebox setting. The original model is constructed via finetuning a pre-trained model with initial \fp parameters trained on the VGGFace2 dataset~\cite{vggface2}. We construct the QAT model by applying TensorFlow tfmot’s \texttt{quantize\_model}~\cite{quantize-model} on the original model. During QAT, we further \finetune this model using the same dataset. Finally, we convert the QAT model to a real \q int8 model with Tflite in order to evaluate it on a resource-constrained device with AArch64 CPU, which would represent the type of processor that might run on an edge device, such as an iPhone~\cite{arm-iphone}. Since Tflite supports only inference and does not expose the gradients, we use QAT's gradients in constructing the \name attacks.

%\paragraph{Implementation.}

%\begin{table}[!t]
%\centering
%\small
%\begin{tabular}{|l|r|r|}
%\hline
%\vtop{\hbox{\strut Model}} & \vtop{\hbox{\strut Accuracy}} & %\vtop{\hbox{\strut Instability (Full-precision)}}\\
%\hline
%\hline
%Full-precision & 99.4\% & -- \\
%\hline 
%QAT & 99.0\% & 0.7\% \\
%\hline
%tflite & 99.0\% & 0.7\% \\
%\hline
%\end{tabular}
%\caption{Comparing the accuracy of the \fp model, the \q model (QAT with int8) and the Tflite model.}
%\label{tab:avg-accuracy-2}
%\end{table}

% Asaf: I'm not sure why the machine in this section matters since we aren't discussing how long it takes to generate the attack etc. Removing this
%\paragraphb{Machine. } Attacks are conducted on a server with four AMD 32-Core EPYC 7502 CPUs, 251 GB of RAM, and four NVIDIA GeForce RTX 3090 Ti GPUs with 24~GB of memory. %Evaluations of the numbers of successful attacks on the \q model is conducted on a cloudlab server with eight 64-bit ARMv8 (Atlas/A57) cores CPUs, 62GiB of RAM.

\begin{figure}[t!]
\captionsetup[subfigure]{justification=centering}
\small
\centering
\begin{subfigure}{0.32\columnwidth}
\includegraphics[width=\textwidth]{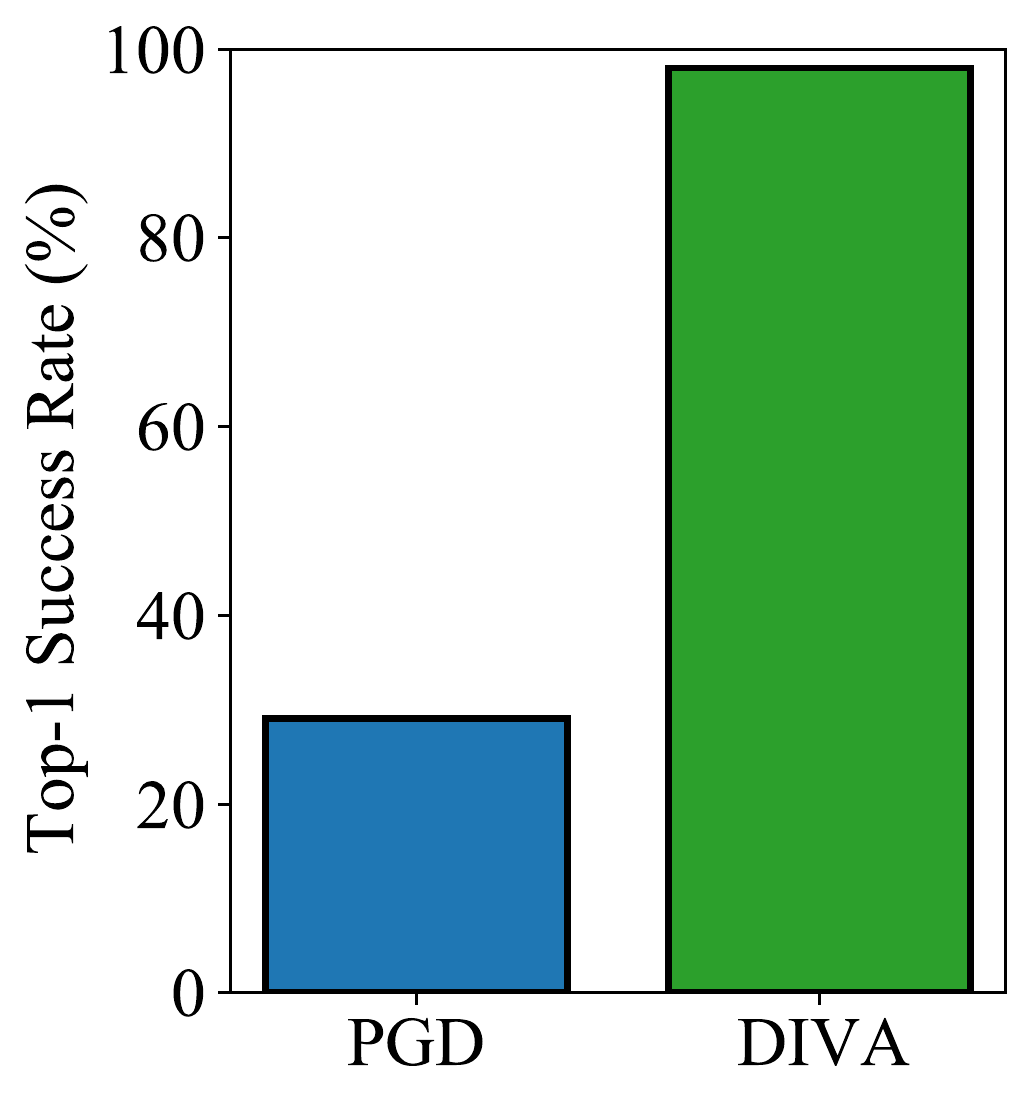}
\caption{Top-1 \protect\\ success rate.}
\label{fig:top-1-attack-FR}
\end{subfigure}~
\begin{subfigure}{0.32\columnwidth}
\includegraphics[width=\textwidth]{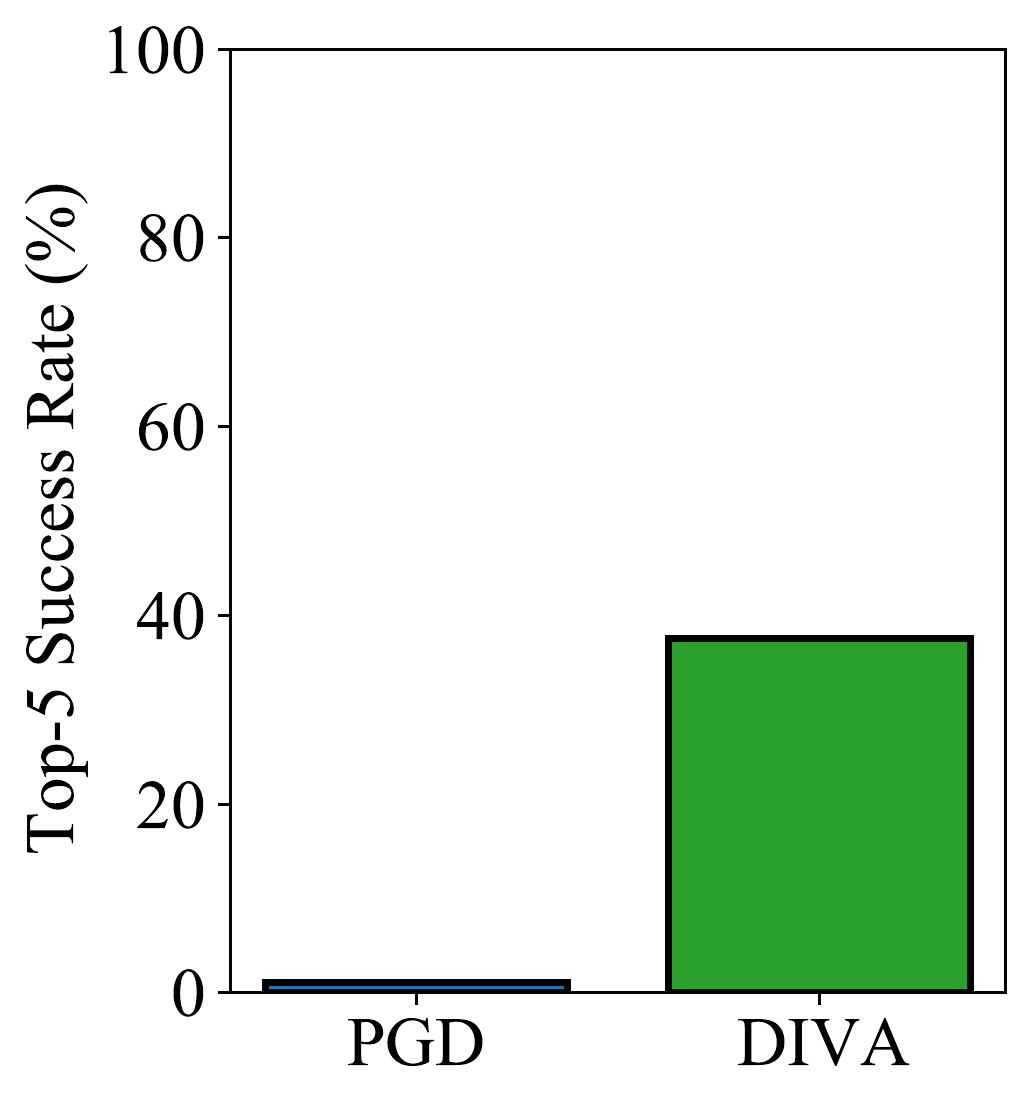}
\caption{Top-5 \protect\\ success rate.}
\label{fig:top-5-attack-FR}
\end{subfigure}~
\begin{subfigure}{0.32\columnwidth}
\includegraphics[width=\textwidth]{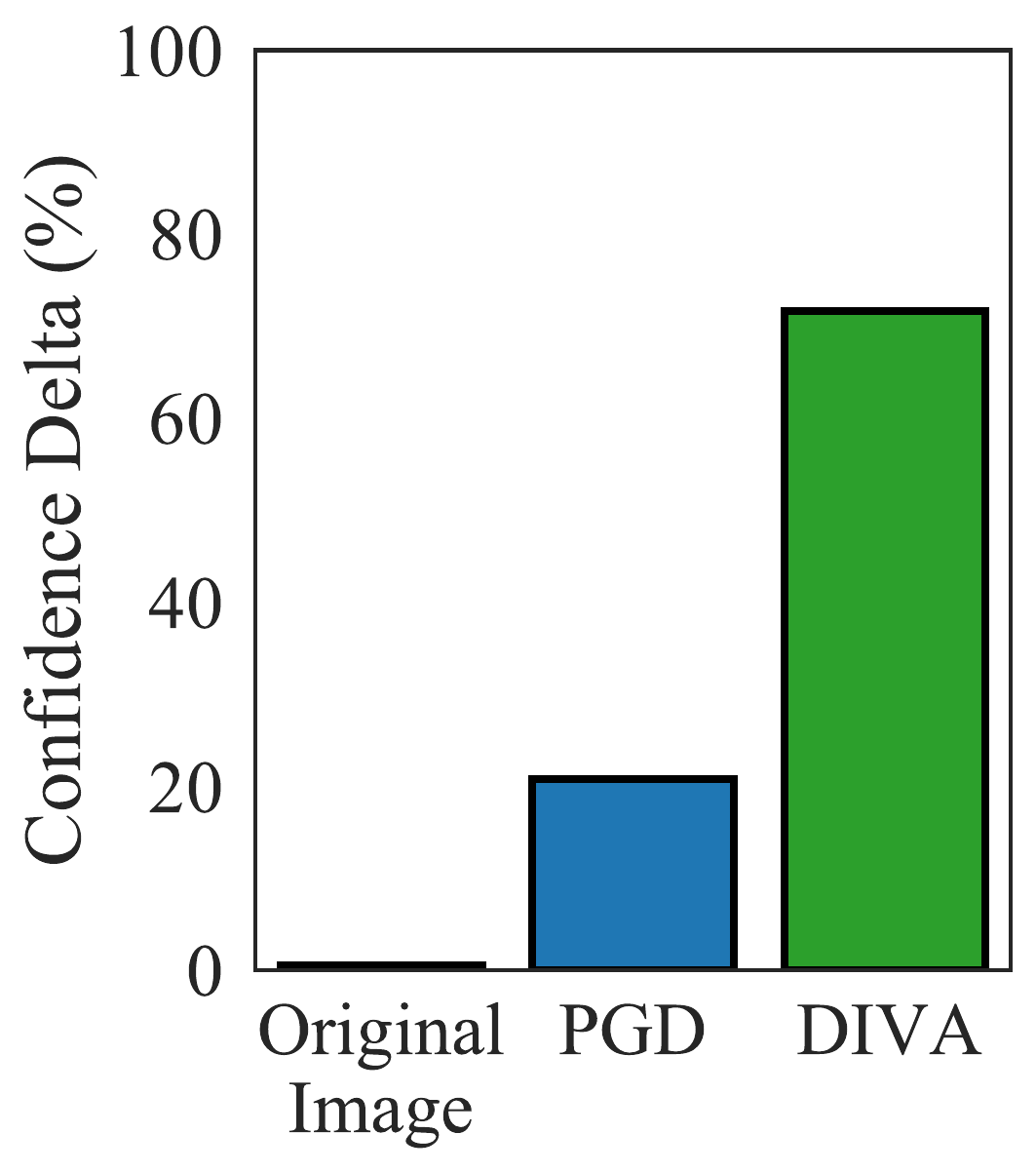}
\caption{Confidence delta. }
\label{fig:confidence-delta-FR}
\end{subfigure}
\vspace{-1em}
\caption{Attacks on the face recognition model.}
\vspace{-1.5em}
\label{fig:top-n-comparison-FR}
\end{figure}

\paragraphb{Evaluation.}
Figure~\ref{fig:top-n-comparison-FR} presents the results of the experiment. On the surface, the \orig and \q models have very similar accuracy: the \orig model has an accuracy of 99.4\%, while the \q model has an accuracy of 99.0\%. However, the high success rates of \name indicate that these models have many subtle differences. Similar to our results on ImageNet, \name significantly outperforms PGD with the face recognition task. \name is somewhat less successful in the top-5 metric compared to the ImageNet models, likely due to the fact that PubFig dataset contains only 150 classes (\ie people), compared to the 1,000-class ImageNet dataset.

\paragraphb{Targeted attack.} We construct a simple targeted attack that attempts not only to fool the adapted model but also lead it to predict a particular person or set of people. This entails adding an additional loss term to our attack, which increases the loss based on its distance away from a one-hot vector with the value of 1 being at the position of the target class. We evaluated the attack on 10 people and were able to target the misclassification on average to a set of 8.3 people (out of the 150 people in the dataset). We believe this attack can be further fine-tuned and be made more accurate.

% Interestingly, even though the instability between Tflite model and the QAT mode is 0\% on the clean validation dataset, it does not mean that these two models are identical when attacked by \name, as \textcolor{red}{447 and 147} of attacks from \name and  PGD succeed on the QAT model while \newtodo{8 and 1} of them do not succeed on the Tflite model. However, since the number of these images is so low, we can conclude that the Tflite model is nearly-equivalent to the QAT model, which is the \q model that would possibly be run on security cameras. 
\section{Other Related Work}

We covered the closely related work in \S\ref{sec:background}. In this section, we discuss more tangentially related work. % on model quantization and the relationship between model compression and robustness.

\paragraphb{Quantization.} Quantization of DNNs is a widely-used technique, in order to allow for efficient inference and to fit larger models on resource constrained edge-device hardware~\cite{JacobKCZTHAK18,DBLP:conf/cvpr/CaiHSV17,DBLP:journals/corr/HanMD15,DBLP:conf/icml/LinTA16}. This work mostly focuses on minimizing the accuracy degradation when compressing a model. Prior work has also shown that even when accuracy is maintained between source model and the quantized version, there is still a prediction divergence between them that can cause instability~\cite{cidon2021characterizing}. This forms the basis for \name's differential approach.

\paragraphb{Model compression and robustness.} Past work has studied the relationship between model compression and robustness~\cite{defensive_q,Ye_2019_ICCV,DBLP:conf/nips/GuiWYYW019,stutz2021bit,liebenwein2021lost}. %Lin \etal~\cite{defensive_q} show that quantized models tend to have reduced robustness relative to the base full-precision model, despite the assumption that quantizing model has a regularization effect. They propose a technique to quantize a model while increasing its robustness. Similarly, 
%Stutz \etal~\cite{stutz2021bit} propose quantization techniques to maintain model robustness to bit errors that happen in low-power devices. Past work has shown that other compression techniques such as pruning~\cite{liebenwein2021lost, Ye_2019_ICCV,DBLP:conf/nips/GuiWYYW019} have a negative effect on model robustness. Gui \etal~\cite{DBLP:conf/nips/GuiWYYW019} and Ye \etal~\cite{Ye_2019_ICCV} both introduce techniques to compress models while maintaining robustness. 
To the best of our knowledge, all past work focused on robustness to traditional adversarial attacks and random noise. Unlike past work, \name %and defense methods% 
focuses on the prediction \emph{divergence} between the original model and its quantized version.

%\paragraph{Adverserial examples.} Generating adversarial examples in deep learning is a heavily researched topic in the ML and security communities~\cite{shan2020gotta,blind-adverserial,SLAP,Waveguard,policy-training,metric:neurips19,gu2014towards,goodfellow2014explaining,carlini2017towards,distillation-defense}. Unlike past adversarial attacks, our attack method specifically targets the divergence between a floating-point model and its quantized version. 

%\paragraph{Other related techniques.} Our attack techniques are related to DeepXplore~\cite{pei2017deepxplore}, a framework for debugging DL models, in which adversarial examples are generated by maximizing the divergence between two models. Our defense method uses Quantized Aware Training~\cite{JacobKCZTHAK18} in which quantization error is introduced while training the model. Model distillation~\cite{BaC14,PolinoPA18} has been shown by past work to help improve the robustness of the student model~\cite{distillation-defense}. We use this insight as part of our defense method.   
\section{Conclusions}

The deployment of DL models in large-scale settings on tens of millions of edge devices creates new security vulnerabilities. This paper highlights a new differential attack, \name, which exploits the subtle differences between the edge-adapted versions of the models and the original server-based model. \name constructs adversarial noise that maximizes the loss of the edge model while minimally affecting the inference of the original model. This causes the edge model to mispredict while significantly increasing the cost of detection and debugging. We adapted this attack to a setting where the attacker only has access to the edge model but not the original model, and showed that it remains effective. 
% We also evaluated several existing and novel defense mechanisms, all of which were unsuccessful in stopping \name, demonstrating that such differential attacks will be hard to defend against.
%While in this paper we primarily focused on model quantization, our approach can be extended to other edge model adaptations, such as compression and pruning. Finally, 
We hope this work opens the door to a new line of research on attacks and defenses that target the variations in models deployed in production.
\section{Acknowledgments}
We thank the reviewers for their comments. This work was supported by a grant from the Columbia Center of AI Technology in collaboration with Amazon, ONR grants N00014-16-12263 and N00014-17-1-2788; NSF grants CNS-2104292, CNS-1750558 and CNS-1564055; a Facebook gift; a JP Morgan Faculty Research Award; and a DiDi Faculty Research Award.
\newpage

\balance
{\footnotesize \bibliographystyle{mlsys2021}
\bibliography{bib}}

\appendix
%%%%%%%%%%%%%%%%%%%%%%%%%%%%%%%%%%%%%%%%%%%%%%%%%%%%%%%%%%%%%%%%%%%%%

%%%%%%%%%%%%%%%%%%%%%%%%%%%%%%%%%%%%

% \begin{figure*}[htp!]
% \small
% \centering
% \begin{subfigure}{0.45\textwidth}
% \includegraphics[width=\textwidth]{figures/Top-1-blackbox.pdf}
% \caption{Attacks on quantized models.}
% \label{fig:blackbox}
% \end{subfigure}~
% \begin{subfigure}{0.45\textwidth}
% \includegraphics[width=\textwidth]{figures/Net_vs_successrate.pdf}
% \caption{Other attacks on quantized models}
% \label{fig:otherbaselines}
% \end{subfigure}

% \caption{\textbf{Additional Attacks  Evaluation in the Quantization Setting}}
% %\vspace{-1em}
% \label{fig:appendix-figs}
% \end{figure*}
\newpage
\section{Artifact Appendix}
\subsection{Abstract}

This artifact appendix helps readers to reproduce the main
experimental results in this paper. In this artifact evaluation,
we show (1) how to create dataset and models for evasive attack evaluation.
(2) how to perform DIVA in white-box, semi-\blackbox and \blackbox setting and perform PGD, Momentum PGD and CW attack as baselines. (3) how to perform analysis on the attack results.

\subsection{Artifact check-list (meta-information)}

\small
\begin{itemize}
  \item {\bf Model:} ResNet50, DenseNet121, MobileNet, VGGFACE\\
  
  \item {\bf Data set:} ImageNet2012, PubFig, MNIST\\

  \item {\bf Run-time environment:} Debian GNU/Linux (or equivalent)\\

  \item {\bf Hardware:} see \S\ref{appendx:hardware}\\

  \item {\bf Metrics:} top-1, top-5 success rate, confidence score delta\\

  \item {\bf How much disk space required (approximately)?:} 15MB (code only)\\

  \item {\bf How much time is needed to prepare workflow using single GPU (approximately)?:} Model generation: 24hrs for 15 models that needs training; Dataset: 6hrs\\

  \item {\bf How much time is needed to complete experiments using single GPU (approximately)?:} Attack: 3.5hrs each in pruning setting, 10 hrs each in quantization setting; evaluation for one attack: 10 hrs\\

  \item {\bf Publicly available?:} Yes\\

  \item {\bf Code licenses (if publicly available)?:} MIT\\

  \item {\bf Workflow framework used?:} see \S\ref{appendx:software}\\

  \item {\bf Archived (provide DOI)?:} 10.5281/zenodo.6084154\\
  
\end{itemize}

%%%%%%%%%%%%%%%%%%%%%%%%%%%%%%%%%%%%%%%%%%%%%%%%%%%%%%%%%%%%%%%%%%%%%
\subsection{Description}

\subsubsection{How to access}

\begin{itemize}
    \item Github repository (\url{https://github.com/WeiHao97/DIVA}) contains latest codes to reproduce all experiment results
    \item Zip file ( \url{https://drive.google.com/file/d/1jy7AVFU8v8lbt8rcTWabV5-WLg9BjRcd/view}) contains code, weights and datasets that can be run independently to reproduce the results from the case study (section 6) in this paper.
\end{itemize}

\subsubsection{Hardware dependencies}
\label{appendx:hardware}

\begin{itemize}
    \item  All experiments on 'server' are conducted on a server with four Intel 20-core Xeon 6230 CPUs, 376 GB of RAM, and eight Nvidia GeForce RTX 2080 Ti GPUs each with 11 GB memory.

    \item  All experiments on 'edge' are conducted on a cloudlab m400 machine with eight 64-bit ARMv8 (Atlas/A57) cores CPUs, 62GiB of RAM. The machine's profile is 'ubuntu18-arm64-retrowrite-CI-2' running on node ms0633 in Utah (\url{https://www.cloudlab.us/}).
\end{itemize}

\subsubsection{Software dependencies}
\label{appendx:software}
\begin{itemize}
\item On server: jupyter notebook, numpy 1.19.5, tensorflow 2.4.1, keras 2.4.3, tensorflow\-model\-optimization, keras\-vggface , matplotlib, livelossplot, spicy, PIL, tensorflow\_datasets, scikit-learn, seaborn, pandas,torch, torchvision, GPUtil, dill, tensorboardX, tables, dssim

\item On edge: jupyter notebook, tflite-runtime

\end{itemize}
\subsubsection{Datasets}

We employ ImageNet, MNIST and PubFig in our experiments. PubFig is included in the zip file. ImageNet2012 has to be download manually from https://image-net.org/challenges/LSVRC/2012/, the code parses it automatically. MNIST is automatically loaded from TensorFlow Datasets by the code.

%%%%%%%%%%%%%%%%%%%%%%%%%%%%%%%%%%%%%%%%%%%%%%%%%%%%%%%%%%%%%%%%%%%%%
\subsection{Installation}

All software dependencies can be installed using pip. \\ dssim package for image similarity analysis can be downloaded from: \url{https://github.com/kornelski/dssim}

%%%%%%%%%%%%%%%%%%%%%%%%%%%%%%%%%%%%%%%%%%%%%%%%%%%%%%%%%%%%%%%%%%%%%
\subsection{Experiment workflow}

\subsubsection{Quantization Experiments (\S\ref{sec:eval-whitebox}--\S\ref{sec:otherbaselines})}
\label{appendx:quantization}
First, generate \orig \fp, quantized and surrogate \fp, quantized models by following the workflow in quantization/model\_generate{*}.ipynb. Next, generate the 3000 images datasets for attack evaluation by following quantization/generateImagePerClass.ipynb. Then, generate the DIVA and baseline attacks by running quantization/{*}.py. See quantization/README.md for more details.

\subsubsection{Pruning Experiments (\S\ref{sec:pruning})}
First, generate pruned models by following the workflow in pruning/ModelGen.ipynb. Next, generate the 3000 images dataset for attack evaluation by following pruning/generateImagePerClass.ipynb. Notice that this dataset is different from the datasets in \S\ref{appendx:quantization}. Then, run attacking scripts under pruning/attacks. See pruning/README.md for more details.

\subsubsection{Case Study Experiments (\S\ref{sec:casestudy})}
Extract the zip file which contains code, weights, and data that can be run independently to reproduce the result of this section.
\begin{itemize}
    \item For untargeted attack: First, split PubFig dataset for model training. Next, construct full-precision, QAT and tflite models and create the attacking dataset following  FR\_server.ipynb on the server. Then generate attacks following untargetted/{*}.py scripts the server and load the results on the edge. On the edge, run FR\_edge.ipynb to evaluate the final top-1/top-5 success rate and the confidence delta on the generated results.
    \item For targeted attack, run targetted.py to generate a dictionary where its key is the person's name and its value is an array of successful attacks.
    
\end{itemize}
See DIVA/quantization/PubFig/README.md for more details

\subsubsection{Attacks under Robust Model Experiments (\S\ref{sec:robust})}
\label{appendx:robust}
First, load the robust-trained full-precision model from \url{https://github.com/MadryLab/robustness} and create the robust-trained quantized model. Then, generate the attack using PGD and DIVA functions on ImageNet Dataset. The whole workflow including evaluation is included in robustness/notebooks/DIVA\_under\_robust\_trained model.ipynb.\\See robustness/README.md for more details.

\subsubsection{MNIST Experiments (Fig \ref{fig:pca-analysis})}
First, train model on the MNIST dataset by following the workflow in quantization/Mnist/ModelGen.ipynb. Next, run DIVA attack by following the workflow in quantization/Mnist/attacks.ipynb. Last, generate visualization for attack results by running quantization/Mnist/PCA\_TSNE.ipynb. See quantization/Mnist/README.md for more details.

%%%%%%%%%%%%%%%%%%%%%%%%%%%%%%%%%%%%%%%%%%%%%%%%%%%%%%%%%%%%%%%%%%%%%
\subsection{Evaluation and expected result} 

The top-1/top-5 success rates for section (\S\ref{appendx:quantization}--\ref{appendx:robust}) can be found both in stdout and in the evaluation script for each experiment. The confidence deltas, DSSIM and evasion cost analysis are calculated in each evaluation script if they are evaluated in the paper.\\
The top-1/top-5 success rates for \S\ref{appendx:robust} can be found in stdout with the format: 

\begin{itemize}
    \item Total: \{\}  Success: \{\}  Q\_W:\{\}  FP\_W:\{\}  Robust\_acc: \{\}
\end{itemize}
success/total gives the success rate and Robust\_acc gives the robustness accuracy evaluated in the paper. Q\_W and FP\_W gives the number of mis-predictions by the full-precision and the quantized model after attack.

%%%%%%%%%%%%%%%%%%%%%%%%%%%%%%%%%%%%%%%%%%%%%%%%%%%%%%%%%%%%%%%%%%%%%
\subsection{Experiment customization}
Hyper-parameters can be customized:
\begin{itemize}
\item $c$: balancing the effect of attack on full-precision and quantized model.  
\item $grad\_iterations$: number of attack steps
\item $step$: step size of each attack step
\item $epsilon$: bound for the attack
\item number of training steps during model generation. 
\end{itemize}

%%%%%%%%%%%%%%%%%%%%%%%%%%%%%%%%%%%%%%%%%%%%%%%%%%%%%%%%%%%%%%%%%%%%%
\subsection{Methodology}

Submission, reviewing and badging methodology:

\begin{itemize}
  \item \url{http://cTuning.org/ae/submission-20200102.html}
  \item \url{http://cTuning.org/ae/reviewing-20200102.html}
  \item \url{https://www.acm.org/publications/policies/artifact-review-badging}
\end{itemize}

\end{document}